\documentclass[10pt,journal,compsoc]{IEEEtran}

\usepackage{amsmath}
\usepackage{amssymb}
\usepackage{graphicx}
\usepackage{algorithm}
\usepackage{algpseudocode}
\usepackage{physics}
\usepackage{braket}
\usepackage{tikz}
\usepackage{subfig}
\usepackage{xcolor}
\usepackage{comment}
\usepackage{orcidlink}
\usepackage{hyperref}
\hypersetup{
    colorlinks=false,
    linkcolor=blue,
    filecolor=magenta,
    urlcolor=cyan,
}

\begin{document}

\title{Hybrid Quantum-Classical Generative Adversarial Networks with Transfer Learning}

\author{
  Asma~Al-Othni\orcidlink{0000-0003-1248-0403}\thanks{Asma~Al-Othni is with the Qatar Center for Quantum Computing, College of Science and Engineering, Hamad Bin Khalifa University, Doha, Qatar. E-mail: \texttt{asal68497@hbku.edu.qa}}, Saif~Al-Kuwari\orcidlink{0000-0002-4402-7710}\thanks{Saif Al-Kuwari is with the Qatar Center for Quantum Computing, College of Science and Engineering, Hamad Bin Khalifa University, Doha, Qatar. E-mail: \texttt{smalkuwari@hbku.edu.qa}}, Mohammad~Mahdi~Nasiri~Fatmehsari\orcidlink{0009-0007-9112-2298}\thanks{Mohammad Mahdi Nasiri Fatmehsari is with the Pasargad Institute for Advanced Innovative Solutions (PIAIS), Tehran, Iran. E-mail: \texttt{mehdi.nafa1373@gmail.com}}, Kamila~Zaman\orcidlink{0009-0003-7151-3700}\thanks{Kamila Zaman is the Founder and Lead Researcher at Sylerk, Islamabad, Pakistan. E-mail: \texttt{kamila.zaman@sylerk.com}}, Ebrahim~Ardeshir-Larijani\orcidlink{0000-0001-6432-2437}\thanks{Ebrahim Ardeshir-Larijani is with the Department of Mathematics and Computer Science, Iran University of Science and Technology, Tehran, Iran.  E-mail: \texttt{larijani@iust.ac.ir}}
}

\maketitle

\begin{abstract}
Generative Adversarial Networks (GANs) have demonstrated immense potential in synthesizing diverse and high-fidelity images. However, critical questions remain unanswered regarding how quantum principles might best enhance their representational and computational capacity. In this paper, we investigate hybrid quantum-classical GAN architectures supplemented by transfer learning to systematically examine whether incorporating Variational Quantum Circuits (VQCs) into the generator, the discriminator, or both improves performance over a fully classical baseline. Our findings indicate that fully hybrid models, which incorporate VQCs in both the generator and the discriminator, produce images with higher quality and achieve more favorable quantitative metrics compared to their fully classical counterparts. In particular, placing the quantum block in the generator appears to accelerate the early emergence of visual structure, whereas placing it in the discriminator slows early visual convergence but improves the final quantitative quality metric. Incorporating quantum blocks into both networks yields the strongest overall performance. Moreover, the model sustains comparable performance even when the dataset size is reduced. Overall, the results underscore that carefully integrating quantum computing with classical adversarial training and pretrained feature extraction can enrich GAN-based image synthesis. These insights open avenues for future work on higher-resolution tasks, alternative quantum circuit designs, and experimentation with emerging quantum hardware.
\end{abstract}

%\begin{IEEEkeywords}
%Generative Adversarial Networks (GANs), Quantum Machine Learning (QML), Hybrid Models, Transfer Learning, Variational Quantum Circuits (VQCs), Image Synthesis.
%\end{IEEEkeywords}

\IEEEpeerreviewmaketitle

\section{Introduction}\label{introduction}
\IEEEPARstart{I}{n} recent years, Artificial Intelligence (AI) has advanced significantly in tasks such as image classification, language translation, and strategic gameplay. Among the different architectures pursued to tackle these tasks is generative models, which have proven especially impactful, enabling the creation of new synthetic data that can mimic or exceed real-world patterns. This capability has far-reaching implications for data augmentation and content creation, positioning generative models as a pivotal aspect of modern AI research~\cite{VAE-QWGAN-2024, ma2025highresqgan, shu2024vqgan}.

Generative Adversarial Networks (GANs) have rapidly evolved to become one of the most powerful frameworks for generative modeling, enabling the synthesis of realistic images and other high-dimensional data. Despite their remarkable achievements, GANs often struggle with issues such as training instability, mode collapse, and the need for large datasets to achieve state-of-the-art results \cite{iglesias2023}. Addressing these challenges has spurred research into new architectures and training strategies, from sophisticated loss functions to transfer learning approaches that leverage pretrained models \cite{arjovsky2017wasserstein, zhao2020, LV-TL}. 
Concurrently, quantum computing has made notable strides,  leading to the emergence of Quantum Machine Learning (QML). QML seeks to harness quantum phenomena, such as superposition and entanglement, to enrich or accelerate classical machine learning workflows. By mapping data onto quantum states and leveraging quantum computing techniques such as parameterized quantum gates, researchers aim to achieve higher expressive power or improved optimization dynamics compared to purely classical methods \cite{Ngo2023}. One key approach involves the use of Variational Quantum Circuits (VQCs), where trainable quantum gates can transform data in ways that may surpass classical transformations, potentially enhancing tasks from classification to generative modeling. Nevertheless, the current Noisy Intermediate-Scale Quantum (NISQ) era imposes limitations such as device noise and limited qubit counts, motivating hybrid quantum-classical techniques that attempt to achieve quantum advantages, taking into account the practical hardware constraints \cite{bharti2022nisq, callison2022hybrid}.

In this paper, we investigate the integration of VQCs into the GAN framework, adopting a hybrid quantum-classical approach while incorporating transfer learning. We take an architectural perspective and ask where a quantum layer should be placed within a GAN in order to be most effective. To this end, we fix a lightweight VQC design and systematically compare fully classical and hybrid GANs in which the same VQC is inserted into the generator, the discriminator, or both. This yields a placement-centric study, in which changes in performance and stability can be attributed to the location of the quantum block rather than circuit depth or qubit count. We conduct experiments on various configurations, and including multiple classes of CIFAR-10 dataset along with extended experiments featuring different sample sizes. We use quantitative metrics, including the Fréchet Inception Distance (FID), Kernel Inception Distance (KID), and Inception Score (IS), and visual inspections to provide a comprehensive assessment of each model's performance.

A key component of our approach is transfer learning, where a pre-trained ResNet-18 backbone is adapted for the discriminator to improve feature extraction. This strategy helps stabilize adversarial training, particularly under limited dataset conditions, while also allowing us to assess how quantum blocks interact with high-quality, pre-trained features. This also provides a strong and consistent baseline for interpreting the contribution of the VQC across the comparative experiments. The results indicate that the fully hybrid model tends to outperform other configurations, exhibiting improved convergence and producing higher-quality images. Furthermore, we show that even when the dataset is halved, the fully hybrid architecture can still achieve outcomes close to those attained with larger datasets, highlighting the resilience and adaptability of our design.

The contributions of this research can be summarized as follows: 
\begin{itemize}
    \item Establish a solid foundation for hybrid quantum-classical GANs integrated with transfer learning.
    \item Provide a placement-centric analysis by fixing a VQC and systematically varying its location (generator, discriminator, or both), isolating the architectural impact of quantum blocks on training dynamics and generative quality.
    \item Investigate how these hybrid quantum-classical methods scale when data is limited or distributed among multiple classes.
\end{itemize}
Beyond these investigations, we provide insights that lay the groundwork for future studies on high-resolution generative tasks and more advanced quantum hardware, suggesting that quantum-enhanced techniques could become a key pillar in next-generation generative modeling.

The rest of this paper is organized as follows.
Section \ref{background} provides background discussions related to GANs, QGANs and Transfer Learning.
Section~\ref{model} details the proposed architecture, including generator, discriminator, variational quantum circuit, and the transfer-learning setup.
Section~\ref{evaluation} explains the experimental configurations, datasets, training setup, and evaluation metrics.  
Section~\ref{binary_results} reports quantitative and qualitative results for the four binary-class experiments, and Section~\ref{multiclass_results} extends the analysis to a multi-class setting.  
Section~\ref{learning_curves} examines learning-curve behavior under reduced sample sizes. Finally, Section~\ref{conclusion} concludes the paper and outlines future research directions. 

\section{Preliminaries and Related Work}\label{background}

\subsection{Generative Adversarial Networks}\label{GANs}%\vspace{-1.0em}
\begin{figure}[ht]
    \centering
    \includegraphics[width=0.5\textwidth]{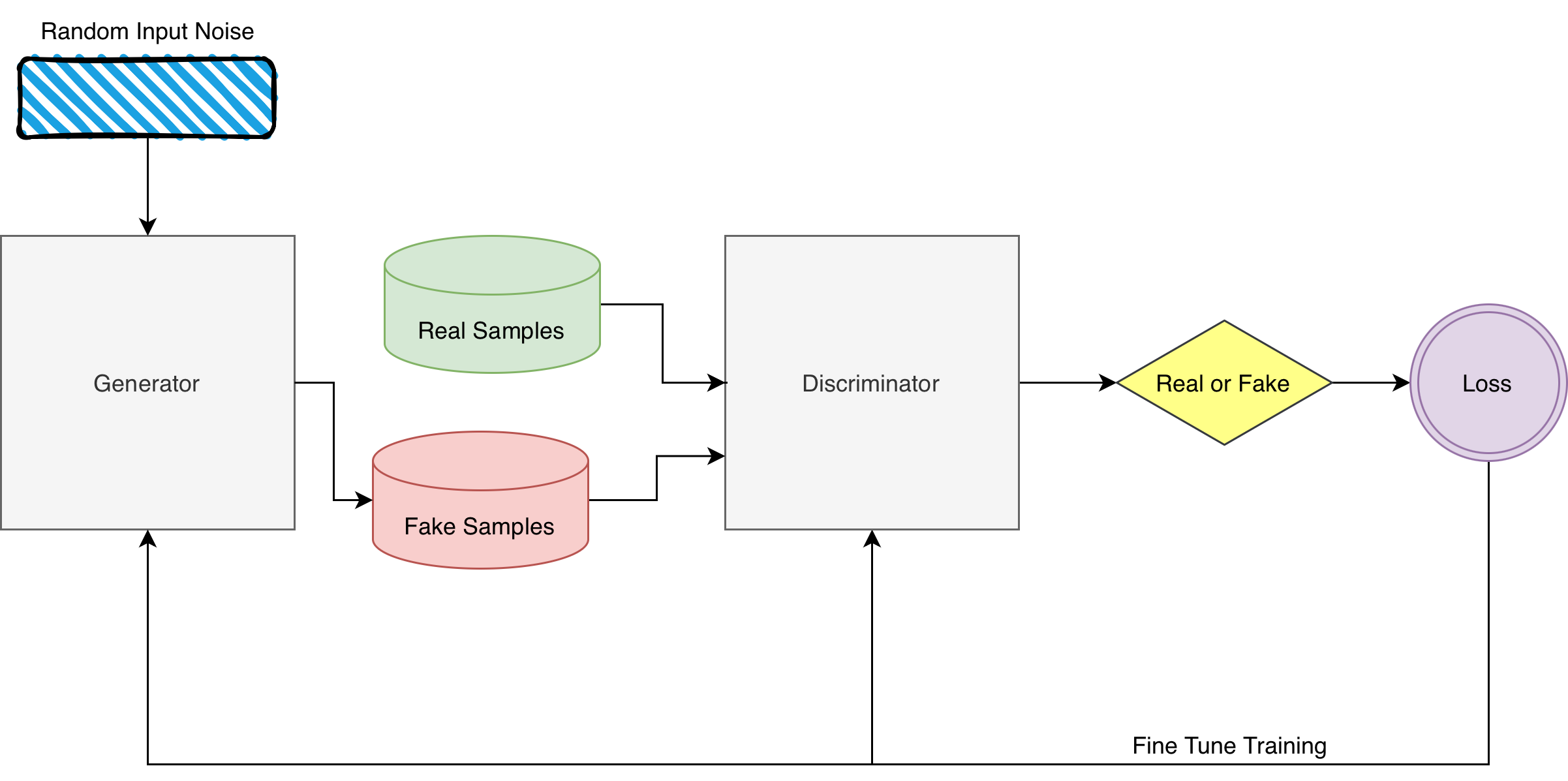}
    \caption{Basic GAN Architecture}
    \label{fig:basic_gan} % Optional: Allows you to reference the figure later using \ref{fig:example}
\end{figure}

Introduced by Goodfellow \emph{et al.}~\cite{goodfellow2014}, GANs are a family of deep learning models consisting of two competing neural networks, a generator and a discriminator, trained together in an adversarial process. The generator network learns to produce synthetic data (e.g., images) that closely mimic the real data distribution, while the discriminator network learns to distinguish between real samples and the generator's synthetic samples. The GAN training procedure is formulated as a minimax (zero-sum) game: the generator strives to minimize its loss (by fooling the discriminator), whereas the discriminator strives to maximize its ability to classify real and fake data correctly. Through this adversarial training, the generator becomes increasingly skilled at creating realistic outputs, while the discriminator becomes more adept at spotting fake inputs until the generated data is almost indistinguishable from real data. Figure \ref{fig:basic_gan} shows how the generator and discriminator interact within the GAN architecture.

GAN formulates the following minimax objective for the two players (generator $G$ and discriminator $D$):
\begin{align*}
    &\mathrm{\min_{G} \max_{D}  V(D,G)} =  \\ &\mathbb{E}_{x \sim p_{\text{data}}}[\log D(x)] + \mathbb{E}_{z \sim p_z}[\log(1-D(G(z)))]
    \label{eq:gan_objective}
\end{align*}

\noindent where $D(x)$ is the discriminator's estimate of the probability that the sample $x$ is real. The generator tries to minimize this value function $V$ (i.e., reducing $\log(1-D(G(z)))$ and making $D(G(z))$ approach 1), while the discriminator tries to maximize it~\cite{goodfellow2014}. 
%\bigskip

In practice, the generator is typically fed a source of random noise (a latent vector) and learns to transform it into realistic data samples. The discriminator is often a binary classifier that produces a probability (real or fake). They are trained in alternating steps: for a batch of data, $D$ is updated to better distinguish real and generated samples, then $G$ is updated to better fool $D$. Over many iterations, this adversarial training ideally reaches a point where the generator’s output distribution matches the real data distribution $p_{g}(x) \approx p_{\text{data}}(x)$ and $D(x) = 0.5$ for all $x$, indicating that the discriminator can no longer tell fake from real.

GANs are typically trained without explicit labels for the generator’s outputs; the only feedback the generator receives is from the discriminator’s judgments. This makes GANs a form of unsupervised (or self-supervised) learning~\cite{eckerli2021gansF}, as the generator learns to model the data distribution without direct ground truth labels for what it generates. The discriminator provides a learning signal to the generator by grading its fakes as ``real” or ``not real”.

%\subsection{GANs Challenges}
Despite the significant progress, training GANs presents several challenges. 
The vanishing gradient problem poses a significant challenge in GAN training. It occurs when gradients become too small during backpropagation, preventing effective weight updates. In GANs, this problem is especially pronounced due to the dynamic interaction between the generator and the discriminator. In particular, it happens when the discriminator learns too quickly, leaving the generator with minimal feedback for improvement. This issue can stall learning and contribute to mode collapse, where the generator produces a limited variation of samples, failing to capture the diversity of the real data distribution~\cite{iglesias2023}. 
Another persistent challenge is the training instability. The adversarial dynamic between the generator and discriminator can lead to oscillations, divergence, or mode collapse during training, necessitating careful hyperparameter tuning and monitoring. Furthermore, GANs often require large and diverse datasets for effective training~\cite{zhao2020}. The generator's ability to produce realistic samples hinges on learning from a rich dataset, which can be a significant limitation when high-quality data is scarce.

\subsection{Quantum Generative Adversarial Networks}\label{QGANs}
Quantum Generative Adversarial Networks (QGANs) leverage quantum computing principles to improve traditional GANs~\cite{Lloyd2018-qgan, hu2019, niu2022}. Early QGAN implementations primarily focused on proof-of-concept models due to the constraints of near-term quantum hardware. The first QGAN models operated on minimal qubit counts, limiting them to generating low-resolution images, often as small as $2\times2$ or $4\times4$ pixels~\cite{pajuhanfard2024qgan}. However, as quantum hardware improves, more sophisticated QGAN architectures have emerged. A significant milestone was achieved by Huang \emph{et al.}~\cite{huang2021experimentalqgan}, who demonstrated an experimental QGAN capable of generating $8\times8$ grayscale handwritten digits using a superconducting quantum processor. Their approach introduced a patch-based quantum generator, where each image was divided into multiple sub-generators, each of which was processed by a dedicated quantum circuit before being reconstructed. Despite utilizing only 5 qubits per patch, QGAN successfully generated recognizable digit images. Subsequent advancements pushed the resolution of QGANs further. Tsang \emph{et al.}~\cite{tsang2022Latent} extended Huang's work by generating full-resolution $28\times28$ images on both MNIST and Fashion-MNIST, marking a significant step forward in quantum generative modeling.

Beyond resolution improvements, researchers have proposed various approaches to incorporate quantum computing into the GAN framework to enhance generative modeling capabilities, leading to a diverse range of QGAN architectures. For example, the QGAN model introduced in~\cite{Lloyd2018-qgan} employs quantum mechanical systems for the discriminator, generator, and data generation process. Similarly, the Quantum Wasserstein GAN (QWGAN) framework~\cite{qwgan2019} integrates quantum circuits to compute the Wasserstein distance between real and generated distributions, a key component of the Wasserstein GAN objective~\cite{arjovsky2017wasserstein}. Quantum Variational Autoencoders (QVAE) have also been explored in generative modeling. One such approach combines a Variational Autoencoder (VAE) with a QWGAN by integrating the VAE’s decoder into the generator of the QGAN~\cite{VAE-QWGAN-2024}. The VAE pre-trains the generator, refining the quality of the latent space representation, which enhances training stability and sample quality. Despite this enhancement, the core training process remains adversarial, relying on the QWGAN framework. These architectural advancements demonstrate the versatility of QGANs in leveraging quantum circuits for more efficient and stable training.

More recently, alternative quantum neural architectures have been introduced to further enhance image fidelity and representation efficiency. One such development is the Quantum Implicit Neural Representation (QINR), proposed in~\cite{ma2025highresqgan}, which integrates implicit neural representations (INRs) with QGANs to improve image quality while reducing the number of trainable parameters. Unlike traditional pixel-based generative approaches, INRs represent data as continuous functions, enabling smoother, higher-resolution image synthesis. By embedding quantum circuits within the INR framework, the QINR-based QGAN exploits quantum superposition and entanglement to capture complex patterns beyond classical generative models. Empirical results indicate that QINR-based QGANs outperform traditional QGANs in sample quality while requiring fewer computational resources, making them a promising direction for further research.

While purely quantum GANs hold promise, their practical implementation remains challenging due to the current limitations of quantum hardware and the complexity of building end-to-end quantum circuits. Hybrid models, which combine classical and quantum components, offer a more realistic and accessible approach to harnessing the advantages of quantum computing. 
A common hybrid approach is to use a classical neural network as the generator or discriminator while integrating a quantum circuit within the other component. Tsang \emph{et al.}~\cite{tsang2022Latent} exemplify this hybrid strategy with a fully quantum generator that slices each $28\times28$ image into 28 column patches, processes every patch through an independent five-qubit variational circuit, combines the outputs together, and forwards the composite image to a classical WGAN-GP critic.  Despite using roughly three orders of magnitude fewer trainable generator parameters than a size-matched classical WGAN, their model achieves comparable visual quality on both MNIST and Fashion-MNIST. Likewise, the authors in~\cite{shu2024vqgan} propose a hybrid GAN architecture in which the generator consists of a quantum variational circuit combined with a one-layer neural network, while the discriminator is a traditional neural network. The authors examined the performance of this quantum-classical GAN numerically by benchmarking it against a classical GAN on the task of handwritten image generation. Their results suggest that the integration of quantum circuits into the GAN framework can potentially improve the training process and the quality of the generated samples.

Beyond image generation, recent quantum machine learning studies have reported task-specific advantages, including hybrid quantum–classical models that outperform classical baselines in ghost imaging with structured one-dimensional data~\cite{Xiao2023GhostImaging}, quantum reinforcement-learning protocols that approach the Heisenberg limit in quantum metrology without critical slowing down~\cite{Xu2025QRLHeisenberg}, and generative quantum advantage for learning and sampling from classically intractable distributions on classical and quantum data~\cite{Huang2025GenerativeQA}.

The progress in QGAN research reflects the broader effort to harness quantum computing for machine learning. While practical, large-scale quantum generative models are still an open challenge, recent developments in QGANs demonstrate tangible improvements in sample quality, convergence stability, and computational efficiency. As quantum hardware continues to advance, QGANs will likely play a pivotal role in next-generation generative modeling, offering unique advantages over purely classical architectures.

\subsection{Transfer Learning}\label{TL}
Transfer Learning (TL) is a machine learning technique in which a model trained on one task is adapted for another related task, effectively enabling the reuse of knowledge between models. 
This approach is particularly beneficial when labeled data for the target task is scarce, as it eliminates the need for training models from scratch, which is both computationally expensive and time-consuming. Pre-trained models, often trained on large-scale datasets, such as ImageNet~\cite{Deng2009ImageNet}, learn hierarchical feature representations that can generalize well to various domains~\cite{HAN-TL}. These learned features can then be either directly used as a fixed feature extractor or fine-tuned to adapt to the specific target task.

Recent research has demonstrated the effectiveness of transfer learning in image classification, showing that models pre-trained on ImageNet tend to generalize well to other classification problems~\cite{Kornblith-TL}. This suggests that feature representations learned from large-scale datasets, like ImageNet, capture essential characteristics of images, making them valuable in a wide range of computer vision tasks~\cite{Kornblith-TL, fang2023doesprogressimagenettransfer}. In~\cite{azizi2021big}, the authors note the common practice of transferring from ImageNet to medical imaging, indicating empirical success despite the inherent differences between natural and medical images. As a result, transfer learning has become a key technique in deep learning, significantly improving efficiency and performance in data-limited scenarios.  

\section{Model}\label{model}
\begin{figure}
  \centering
  \includegraphics[scale=0.3]{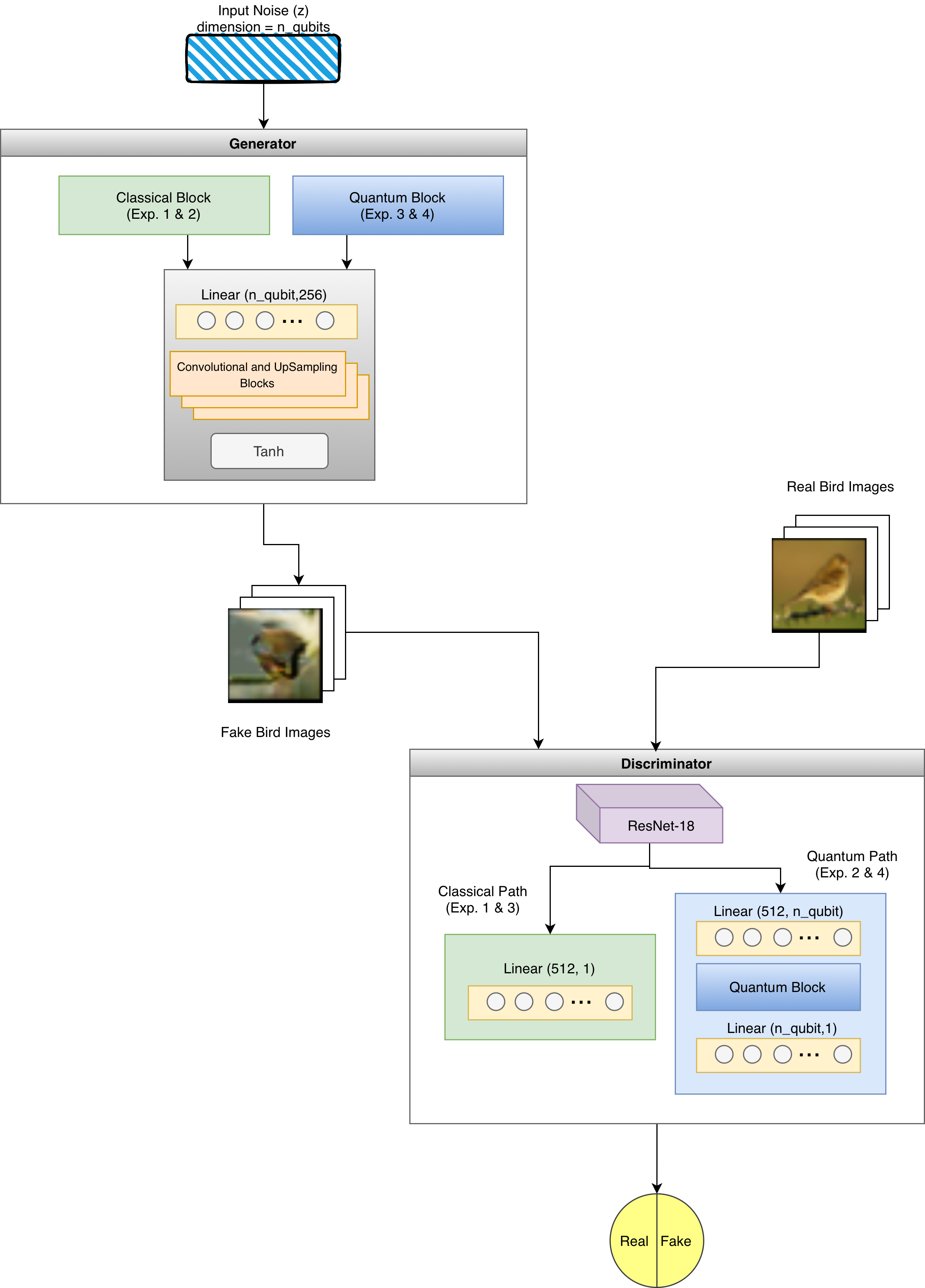}
  \caption{Proposed Model Architecture}
  \label{fig:model_arch} % Optional: Allows you to reference the figure later using \ref{fig:example}
\end{figure}

In this section, we present our proposed QGAN model by describing the generator, the discriminator, the VQC we adopt in both, and the transfer-learning setup. Figure \ref{fig:model_arch} illustrates the proposed model architecture corresponding to the four core configurations, which are evaluated in Section~\ref{evaluation}.

\subsection{Generator}%\vspace{-1.0em}
As illustrated in Figure \ref{fig:model_arch}, the generator has two variants, one with a classical block and one with a quantum block. Apart from this first block, the rest of the architecture of the generator component is identical, ensuring that any observed performance differences arise primarily from the presence or absence of quantum blocks, rather than from broader architectural changes. The generator is structured as follows:
%in each experiment, the generator follows the same structure, with only the first processing block differing: classical for Experiments 1 and 2, quantum for Experiments 3 and 4. The remainder of the architecture is unified, ensuring that any observed performance differences arise primarily from the presence or absence of quantum blocks, rather than from broader architectural changes. The generator is structured as follows: 

\subsubsection{Latent Vector ($z$)} Each training iteration begins with the sampling of a noise vector, where dim($z$)= $n_{\text{qubits}} = 5$. We select five qubits to keep the quantum circuit simulation computationally feasible while still affording sufficient expressiveness in the latent space. This design decision is consistent with the approach taken by~\cite{tsang2022Latent}, who constrain the latent space to low dimensions (less than 10) due to resource limitations inherent in quantum simulations, yet demonstrate that even a low-dimensional latent space can yield smooth and meaningful mappings between latent vectors and generated images.

\subsubsection{Learning Block}
In our architecture, the learning block can either be classical or quantum. The Classical Block and the Quantum Block serve the same purpose (to process the $n_{\text{qubits}}$-dimensional latent vector $\mathbf{z}$), yet they do so using different mechanisms. Both blocks are designed to have a comparable number of \emph{trainable parameters} to ensure a fair baseline for assessing quantum versus classical performance.

\paragraph{Classical}
This sub-network transforms $\mathbf{z} \in \mathbb{R}^{n_{\text{qubits}}}$ through two fully connected layers with a ReLU in between. Concretely, it starts with A linear layer projects $\mathbf{z}$ from $\mathbb{R}^{n_{\text{qubits}}}$ to $\mathbb{R}^{1}$ (with no bias). Next, we have a ReLU activation to introduce nonlinearity. Finally, we add a second linear layer (with bias) that maps $\mathbb{R}^{1}$ back to $\mathbb{R}^{n_{\text{qubits}}}$.
%        \begin{enumerate}
%        \item A linear layer projects $\mathbf{z}$ from $\mathbb{R}^{n_{\text{qubits}}}$ to $\mathbb{R}^{1}$ (with no bias).
%       \item A ReLU activation introduces nonlinearity.
%        \item A second linear layer (with bias) maps $\mathbb{R}^{1}$ back to $\mathbb{R}^{n_{\text{qubits}}}$.
%        \end{enumerate}
Let $n_{\theta}^{(\text{classical})}$ denote the total number of trainable parameters in the classical block. The first linear layer has $n_{\text{qubits}} \times 1$ (weights only, no bias) parameters, while the second linear layer has
\[
  1 \times n_{\text{qubits}} \; (\text{weights})
  \;+\; n_{\text{qubits}} \; (\text{biases})
  \;=\; 2\,n_{\text{qubits}}.
\]
Hence,
\[
  n_{\theta}^{(\text{classical})}
  \;=\;
  (n_{\text{qubits}})
  \;+\;
  (2\,n_{\text{qubits}})
  \;=\;
  3\,n_{\text{qubits}}.
\]
For $n_{\text{qubits}} = 5$, this block thus contains $15$ trainable parameters.

\paragraph{Quantum} In the quantum variant, a VQC is used to encode $\mathbf{z}$  into qubit rotations rather than purely classical linear transformations. Specifically, each component of $\mathbf{z}$ is loaded into the circuit via an $\mathrm{RY}$ rotation on a corresponding qubit. Subsequently, A chain of $\mathrm{CNOT}$ gates introduces entanglement among the qubits. Each qubit then undergoes a set of trainable single-qubit rotations: $\mathrm{RX}(\theta_{i,0})$, $\mathrm{RY}(\theta_{i,1})$, and $\mathrm{RZ}(\theta_{i,2})$. Because each qubit has three trainable parameters $\{\theta_{i,0}, \theta_{i,1}, \theta_{i,2}\}$, the total parameter count for the quantum block is:
\[
  n_{\theta}^{(\text{quantum})}
  \;=\;
  3\,n_{\text{qubits}}.
\]
Thus, for $n_{\text{qubits}} = 5$, the circuit also contains $15$ trainable parameters, which match the parameter budget of the classical block. After applying these rotations and entanglement gates, we measure the expectation value of the Pauli-$Z$ operator on each qubit. This yields an $n_{\text{qubits}}$-dimensional output vector, which replaces the final output of the classical block. 
By constructing both blocks to produce the same dimensional output ($\mathbb{R}^{n_{\text{qubits}}}$) and assigning them comparable parameter counts ($3\,n_{\text{qubits}}$), we ensure a balanced comparison between the quantum and classical approaches for this critical first stage of the generator. The quantum circuit architecture is discussed in detail in Section \ref{sec:QC}.

\subsubsection{Fully-Connected and Batch Normalization} After the latent vector $\mathbf{z}$ has been processed by the classical or quantum block, the resulting $\mathbb{R}^{n_{\text{qubits}}}$-dimensional output is mapped via a fully connected (FC) layer into a larger feature space of $(256 \times 4 \times 4)$. This FC layer expands the intermediate representation from a relatively small dimension (e.g., 5) to $256 \times 4 \times 4 = 4096$ features. A one-dimensional batch normalization layer then normalizes and scales these activations across the batch, leading to more stable training and improved convergence. The resulting vector of 4096 elements is subsequently reshaped into a $4 \times 4$ spatial grid with 256 channels, forming the initial low-resolution feature map upon which further upsampling and convolutional operations are applied.

\subsubsection{Residual Upsampling Blocks} 
\begin{figure}[ht]
          \centering
          \includegraphics[width=0.45\textwidth]{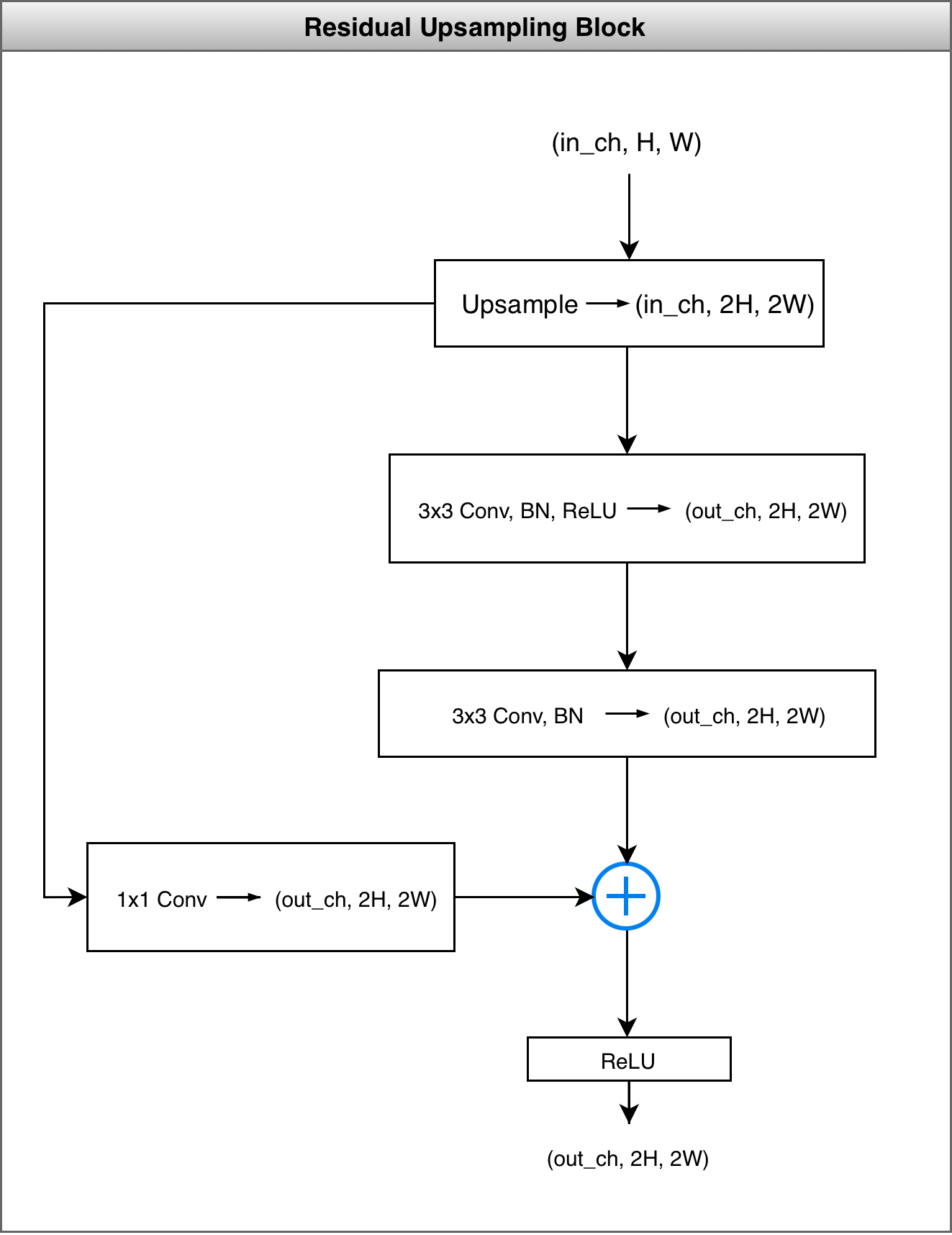}
           \caption{Residual Upsampling Block}
          \label{fig:ResUpBlock} % Optional: Allows you to reference the figure later using 
        \end{figure}

Two sequential Residual Upsampling Blocks gradually increase the spatial resolution of the feature maps. The first block expands the feature maps from $(256,\,4,\,4)$ to $(128,\,8,\,8)$, and the second block further expands them from $(128,\,8,\,8)$ to $(64,\,16,\,16)$. These blocks are designed to progressively increase spatial resolution while preserving crucial information via shortcut (residual) connections. In each block, as iullestrated in Figure \ref{fig:ResUpBlock}), the input feature map is first upsampled using nearest-neighbor interpolation, which doubles its height and width. The upsampled output is then refined through a series of standard convolutional layers (each followed by batch normalization and a ReLU activation). Concurrently, a $1\times1$ convolution is applied to the upsampled input to align the channel dimensions with the output of the convolutional path. This alignment enables an element-wise addition between the two paths, ensuring that the original features are preserved and that the gradient flow remains smooth, thereby mitigating the vanishing or exploding gradients commonly encountered in deeper networks.

This approach is inspired by recent advances in GANs, where a step-by-step increase in resolution, combined with residual connections, has been shown to stabilize training yielding higher-fidelity synthetic images~\cite{Karras2018}. By incrementally building up to the final resolution (e.g., $32\times32$), these modules provide the generator with a robust mechanism to integrate both coarse and fine features, ultimately improving image realism and reducing common training instabilities.
  
\subsubsection{Final Upsampling and Convolution} After the second residual upsampling block, the feature map is further upsampled using nearest-neighbor interpolation, increasing its spatial dimensions from $(64,16,16)$ to $(64,32,32)$. Subsequently, a final $3\times3$ convolution is applied to project the 64 channels down to 3 output channels corresponding to the RGB color channels. This final convolution refines the upsampled features, producing the synthesized image with the desired resolution and color format.
  
\subsubsection{Tanh Output} A hyperbolic tangent (\(\tanh\)) activation function is applied to the final output of the generator, scaling the pixel intensities to the range \([-1, +1]\). This normalization is consistent with the preprocessing of CIFAR-10 images, which are also normalized to \([-1, +1]\).

\subsection{Discriminator} \label{discriminator}%\vspace{-1.0em}
The discriminator likewise has two variants, a fully classical variant and a hybrid variant that appends a quantum block to the final layers. Both start from the ImageNet-pretrained ResNet-18 weights
described in Section~\ref{Probosed_TL}. All input images (originally $32 \times 32$) are resized to $224 \times 224$ to be compatible with the ResNet-18 backbone.

\subsubsection{Fully Classical Discriminator}\leavevmode\par

In the fully classical setting, we leverage the pretrained ResNet-18 backbone for robust feature extraction. To tailor ResNet-18 for our GAN framework and the low-resolution CIFAR-10 images, we make several key modifications:
\begin{enumerate}
    \item The original $7\times7$, stride-2 \textit{conv1} is replaced with a $3\times3$, stride-1 convolution wrapped in \texttt{spectral\_norm} to preserve spatial detail in $32\times32$  images and to stabilize GAN training.

    \item The standard ResNet-18 architecture includes a max-pooling layer early in the network to quickly reduce spatial dimensions. However, since our inputs are low-resolution images (e.g., $32\times32$ images), preserving fine-grained spatial details is critical to differentiate between real and generated images. Removing the max-pooling layer ensures that these details are maintained throughout the network.
    \item The final FC layer of ResNet-18, originally used for classification, is replaced with an identity function, which outputs a 512-dimensional feature vector. This modification retains the rich feature representation for further processing, rather than constraining it to a fixed classification space.
\end{enumerate}
After these modifications, a linear layer maps the 512-dimensional feature vector to a single scalar value for real/fake discrimination.

\subsubsection{Hybrid Discriminator}\leavevmode\par

In the case of a hybrid discriminator, we retain the ResNet-18 backbone with the same adaptations (\textit{conv1} resizing, \textit{max-pool} removal, and identity mapping of the final FC layer) to extract a 512-dimensional feature vector. The subsequent layers are modified as follows:
\begin{enumerate}
    \item Projection to $n_{\text{qubits}}$: A linear layer projects the 512-dimensional feature vector to a lower-dimensional space of size $n_{\text{qubits}}$. This projection aligns the feature vector with the input required by the quantum block.
    \item Quantum Block: The $n_{\text{qubits}}$-dimensional vector is processed by a variational quantum circuit (similar to the one used in the generator). This circuit applies trainable single-qubit rotations and entangling gates, returning a vector of the same dimension.
    \item Final Linear Output: A final linear layer maps the output of the quantum block to a single scalar value for real/fake discrimination.
\end{enumerate}
By integrating a quantum block into the final classification layers in the hybrid configuration of the discriminator, we aim to leverage potential advantages from richer, non-classical decision boundaries.

\subsection{Variational Quantum Circuit}\label{sec:QC}
This integration of quantum circuits into classical neural architectures reflects a broader trend in hybrid quantum-classical learning systems, where quantum layers are ``dressed'' with classical pre-processing or post-processing components. Such configurations, often referred to as \emph{dressed quantum circuits}, have been shown to improve flexibility and training dynamics, especially in scenarios involving transfer learning~\cite{Mari2020transferlearningin}.

Our model deploys these dressed quantum blocks in the generator, discriminator, or both, while maintaining a unified circuit design that balances expressive power with computational feasibility. Figure~\ref{fig:QC} shows the circuit diagram used in this project, which includes: 

    \begin{figure}[ht]
      \centering
      \includegraphics[width=0.5\textwidth]{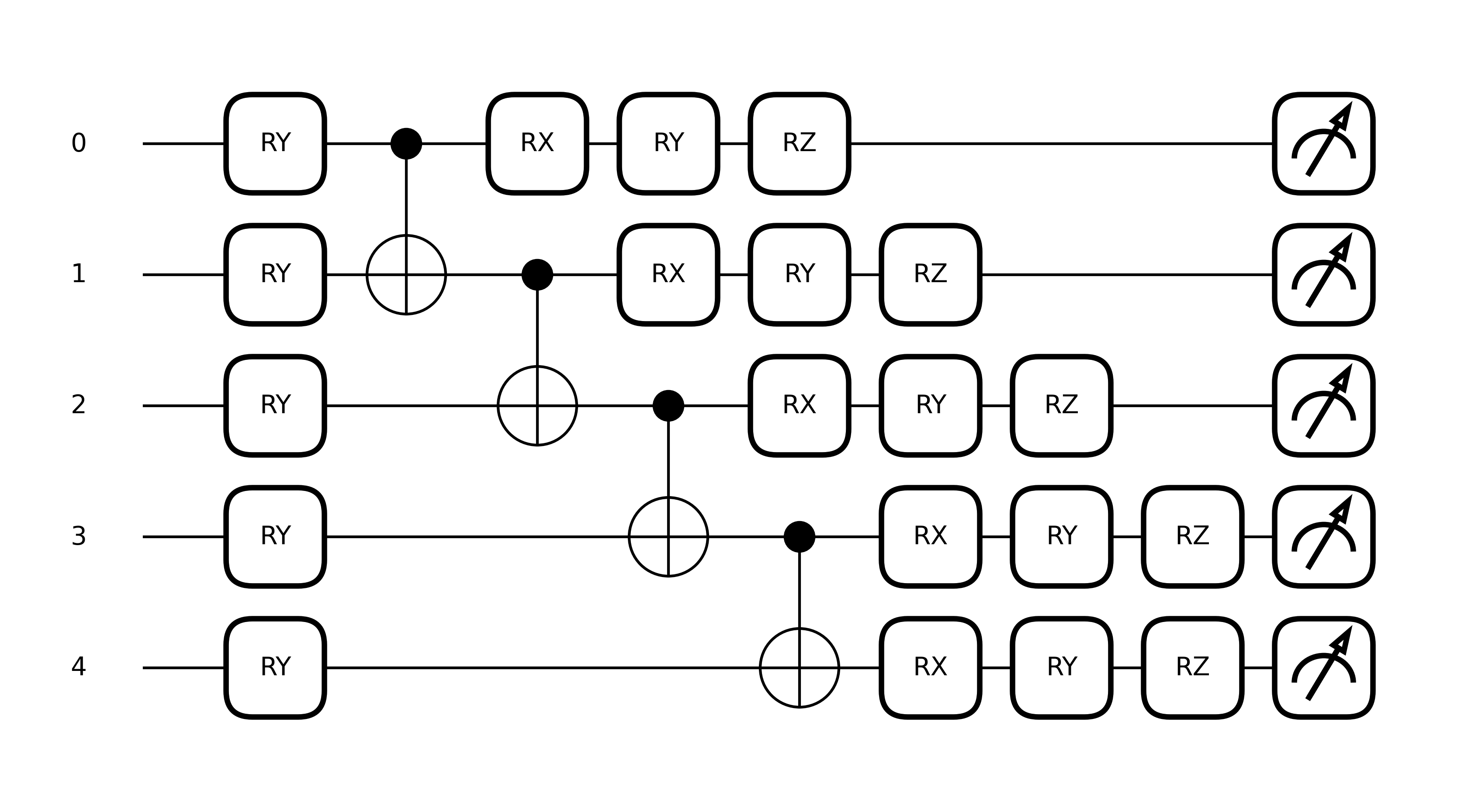}
      \caption{Quantum Circuit Diagram}
      \label{fig:QC} % Optional: Allows you to reference the figure later using \ref{fig:example}
    \end{figure}
    
\begin{enumerate}
    \item \textit{Data Encoding:} Each qubit's rotation angle (\(\mathrm{RY}\) gate) is set to a specific component of the input vector, whether that vector represents latent noise (in the generator) or projected features (in the discriminator). This step embeds classical data into the quantum state.
    \item \textit{Entanglement:} Sequential \(\mathrm{CNOT}\) gates between neighboring qubits introduce correlations that are difficult to replicate in purely classical networks. By tying the qubits' states together, the circuit can learn richer representations, as demonstrated in multiple studies showing that quantum circuits equipped with entangling gates outperform non-entangling variants in generative modeling tasks and QNN architectures. These results highlight that entanglement is a key resource for capturing complex data distributions~\cite{Benedetti2019Generative,PhysRevA.109.032413}.
    \item \textit{Trainable Gates:} Each qubit undergoes parameterized single-qubit rotations \(\mathrm{RX}(\theta_{i,0})\), \(\mathrm{RY}(\theta_{i,1})\), and \(\mathrm{RZ}(\theta_{i,2})\). The circuit thus has \(3 \times n_{\text{qubits}}\) learnable parameters, tuned via backpropagation to improve performance. Keeping \(n_{\text{qubits}}\) relatively small ensures that simulation overhead remains manageable.
    \item \textit{Measurement:} Finally, the circuit measures each qubit in the Pauli-\(Z\) basis, returning a real-valued expectation value per qubit. For a batch of size \(N\), this yields an \((N,\, n_{\text{qubits}})\) output tensor of expectation values, which is then passed to the next stage of the network (e.g., a fully connected layer or the generator’s upsampling blocks).
\end{enumerate}

Overall, VQC implements a single-layer ansatz consisting of data-encoding \(\mathrm{RY}\) rotations, a nearest-neighbor entangling chain, and one block of \(\mathrm{RX}\)–\(\mathrm{RY}\)–\(\mathrm{RZ}\) rotations per qubit, followed by Pauli-\(Z\) measurements. This standardized, shallow design aligns with NISQ-oriented QML practices and allows a controlled comparison across different placements of the quantum layer within the GAN.

During each training iteration, the discriminator is updated by computing a loss from real images (labeled as 1) and images generated by the generator (labeled as 0), and then backpropagating the sum of these losses to update its parameters. Subsequently, the generator is optimized to produce images that the discriminator misclassifies as real. This adversarial training loop forms the backbone of our experimental framework and provides a robust foundation for comparing fully classical and hybrid quantum-classical architectures.

\subsection{Transfer Learning}\label{Probosed_TL}
Compared to the extensive exploration and proven success of transfer learning in discriminative tasks, such as image classification and object detection, its application to generative tasks, particularly image synthesis, has received relatively less attention~\cite{zhao2020}. Recent studies have shown that transfer learning can also be beneficial in improving the performance of GANs in target domains with limited data~\cite{zhao2020, LV-TL}.  A key common insight from these studies is that the low-level filters of a pre-trained GAN discriminator, which capture general visual features like edges and textures, can be effectively leveraged to aid generation in the target domain. By freezing or fine-tuning the low-level layers of the discriminator, and only training the higher-level layers, the generator can be guided to produce more realistic samples, even when the target dataset is small. In contrast, the generator must typically be trained from scratch or carefully fine-tuned, as its task of learning a novel and specific data distribution rarely benefits from previously learned discriminative features. This distinction emphasizes why practitioners commonly apply transfer learning to discriminators to expedite GAN training and enhance image quality, stability, and convergence speed~\cite{Chong2024P2D, Mo2020FreezeD}.

Following the discriminator design in Section~\ref{discriminator}, we apply transfer learning exclusively to the discriminator network. Specifically, we load the ImageNet-pretrained ResNet-18 weights released by He~\emph{et~al.}~\cite{He2016ResNet}, apply our architectural modifications, and fine-tune all layers. The pretraining provides robust edge and texture detectors, while full fine-tuning enables the model to adapt entirely to the CIFAR-10 domain and the adversarial training objective, without the risk of freezing suboptimal filters. The architectural changes described earlier ( \textit{conv1} resizing, removal of the early \textit{max-pool}, and replacement of the final FC layer with \texttt{Identity}) are applied to both discriminator variants prior to fine-tuning. ResNet-18 is selected for its balance between computational efficiency and representational performance, offering a lightweight yet effective backbone for transfer learning on CIFAR-10 within hybrid quantum-classical workflows. In this study, transfer learning serves as a strong pretrained backbone for the discriminator, supporting feature extraction and training stability, while also providing a consistent classical foundation across Experiments 1--4 so that the effect of VQC placement can be assessed more clearly.

\section{Evaluation}\label{evaluation}
 Our work comprises four primary configurations (Experiments~1--4) along with Experiment~5, which is an extension of Experiment~4. The key variations lie in the integration of quantum blocks within the generator and/or the discriminator. Specifically, Experiment~1 employs a fully classical architecture for both the generator and the discriminator; Experiment~2 uses a classical generator with a quantum-enhanced discriminator; Experiment~3 utilizes a quantum-enhanced generator while retaining a fully classical discriminator; and Experiment~4 integrates quantum blocks in both the generator and the discriminator. Experiment~5 further examines the model under more demanding conditions, including multiple classes and a longer training duration. 

To evaluate performance, we employ a comprehensive set of metrics, including adversarial training losses, extended quantitative measures such as FID, KID, and IS, as well as qualitative assessments via visual inspection of generated images. Together, these metrics allow for a detailed analysis of convergence, training stability, and the overall quality of the generated images.

\subsection{Experiments}\label{sec:experiments}
The purpose of this research is to explore the integration of quantum computing elements into GANs, specifically focusing on their impact when combined with classical transfer learning techniques in the discriminator. Although our experiments are based on the CIFAR-10 dataset, we design a progression of increasingly challenging settings. We propose four core experimental setups, each training on a reduced CIFAR-10 dataset (initially filtering only the ``Bird” class). The experiments vary by whether the generator and/or discriminator contain quantum blocks. Finally, in Experiment 5, we extend the fourth experiment to investigate multiple classes (Birds, Cars, Dogs) with longer training schedules (500 epochs) and varying sample sizes to assess scalability and robustness. These multi-class and reduced-sample settings serve as stress tests that push the models beyond simpler regimes. The five core experiments conducted in this study are as follows:
\begin{itemize}
    \item \textbf{Experiment 1 (Baseline, Fully Classical):} A classical generator paired with a classical discriminator utilizing transfer learning. This serves as the reference model for evaluating quantum enhancements.
    \item \textbf{Experiment 2 (Classical Generator, Hybrid Discriminator):} A classical generator is combined with a hybrid discriminator that integrates both transfer learning and a quantum processing block.
    \item \textbf{Experiment 3 (Hybrid Generator, Classical Discriminator):} A quantum-enhanced generator is trained alongside a classical discriminator that incorporates transfer learning.
    \item \textbf{Experiment 4 (Fully Hybrid, Single-Class):} Both the generator and the discriminator employ quantum processing components, alongside transfer learning in the discriminator.
    \item \textbf{Experiment 5 (Fully Hybrid, Multi-Class):} To further assess the robustness and generalizability of the fully hybrid model (Experiment 4), we conducted extended experiments under more demanding conditions. In these extended experiments, the scope was broadened beyond the single ``Bird” class by performing separate runs on three distinct classes: Birds, Cars, and Dogs. Moreover, the training duration was increased from 100 epochs to 500 epochs, thereby enabling a deeper analysis of long-term stability and scalability in a hybrid quantum-classical setting. For evaluation, we compared image quality and training dynamics using two different sample sizes: one experiment utilized 5,000 samples, and another employed 2,500 samples. This comprehensive evaluation framework facilitates a detailed examination of the fully hybrid architecture's performance under diverse data regimes and extended training durations.
\end{itemize}

To isolate the effect of the quantum components, we intentionally avoid using deeper or more sophisticated GAN backbones. More expressive classical architectures could achieve strong performance on their own, making it difficult to detect whether any observed gains are due to the quantum layer or simply the added classical capacity.

\subsection{Dataset}\label{sec:dataset}%\vspace{-1.0em}
All experiments utilize the CIFAR-10 dataset, which contains 60,000 color images of size 32×32 spanning 10 classes. Because the discriminator uses a pretrained ResNet-18 (originally trained on ImageNet), we require 3-channel RGB images to match the network’s expected input format, and CIFAR-10 conveniently provides color images in that shape. Moreover, CIFAR-10’s relatively small resolution and simpler content reduce computational overhead, which is particularly important when simulating quantum circuits, allowing for faster training iterations. 

For the initial 100-epoch runs of each experiment, we filter the dataset to include only the ``Bird” class (index 2), thus isolating a single object category and focusing on the generative capabilities without the multi-class confusion. In Experiment 5, we broaden this scope by including three classes, Birds (index 2), Cars (index 1), and Dogs (index 5), and increasing training to 500 epochs, thus assessing how the fully hybrid quantum-classical architectures scale with multiple classes and longer training schedules.

%\subsection{Data Splitting and Loading}
In our experiments, after filtering CIFAR-10 for the target (Birds) class, 5,000 images from the training set are used for model training. Since the CIFAR-10 test set contains about 10,000 images, filtering it for the target class yields roughly 1,000 images for evaluating generative performance. Using this held-out test subset for evaluation ensures that our performance metrics are computed on images not seen during training. Moreover, all images are loaded in batches of size 8 to accommodate quantum-circuit overhead, and normalized to the range \([-1, +1]\) across each channel.

\subsection{Training Setup}\label{sec:training_setup}%\vspace{-1.0em}
The training setup for our GAN experiments is designed to balance computational efficiency with training stability, particularly given the overhead associated with quantum circuit simulations. For Experiments~1 through 4, each model is trained for 100 epochs to evaluate the convergence and performance differences, while Experiment~5 (incorporating multiple classes of Birds, Cars, and Dogs) is trained for 500 epochs to assess scalability and long-term stability. To maintain a controlled comparison between classical and hybrid configurations, all experiments use the same set of training hyperparameters. In particular, all experiments use a consistent batch size of 8, which ensures that even the hybrid quantum-classical models are trained within the available hardware/simulation constraints. We utilize the Adam optimizer with a learning rate of $2 \times 10^{-4}$ and momentum parameters $\beta = (0.5, 0.999)$, settings that are standard in the GAN literature and effective for preventing mode collapse by smoothing gradient updates. The Binary Cross-Entropy with Logits loss function (BCEWithLogitsLoss) is applied to both the generator and discriminator, allowing the discriminator's output to be interpreted as raw logits while avoiding numerical issues associated with separate sigmoid activations. To ensure robustness for the binary-class configurations (Experiments 1–4), each model was evaluated across four independent runs with different random initializations. 

\subsection{Evaluation Metrics}\label{sec:evaluation_metrices}
We assess the performance of our GAN models using a multifaceted evaluation framework. First, we monitor the generator and discriminator losses, which provide initial insights into the adversarial training dynamics. Second, we perform periodic visual inspections by generating sample images to subjectively assess realism and diversity. Third, we compute extended quantitative metrics to rigorously evaluate the quality of the generated images. These metrics include FID, KID, and IS.

\paragraph*{Fréchet Inception Distance (FID)} FID measures the distance between the feature distributions of real and generated images. It is computed as:
\begin{equation*}
\mathrm{FID} = \|\mu_r - \mu_g\|^2 + \operatorname{Tr}\Bigl(\Sigma_r + \Sigma_g - 2\Bigl(\Sigma_r \Sigma_g\Bigr)^{\frac{1}{2}}\Bigr),
\label{eq:FID}
\end{equation*}
where \(\mu_r\) and \(\Sigma_r\) denote the mean and covariance of the features extracted from real images, and \(\mu_g\) and \(\Sigma_g\) denote those of the generated images. Lower FID values indicate that the two distributions are closer, implying that the generated images more closely resemble the real ones.

\paragraph*{Kernel Inception Distance (KID)} KID is based on the squared Maximum Mean Discrepancy (MMD) using a polynomial kernel, and it provides an alternative measure that is less sensitive to the choice of feature space while offering an unbiased estimate with finite sample sizes. KID is defined as:
\begin{align*} \label{eq:KID}
\mathrm{KID} &= \frac{1}{n(n-1)} \sum_{i \neq j} k(x_i, x_j) \\ &+ \frac{1}{m(m-1)} \sum_{i \neq j} k(y_i, y_j) - \frac{2}{nm} \sum_{i,j} k(x_i, y_j),
\end{align*}
where \(x_i\) and \(y_i\) represent the features of real and generated images respectively, \(n\) and \(m\) are the number of real and generated samples, and \(k(\cdot,\cdot)\) is a polynomial kernel function (e.g., \(k(a,b)=\left(\frac{a^\top b}{d}+1\right)^3\)). Lower KID scores similarly indicate a closer match between the distributions.

\paragraph*{Inception Score (IS)} IS evaluates both the quality and diversity of generated images. It is calculated as:
\begin{equation*}\label{eq:IS}
\mathrm{IS} = \exp\Bigl( \mathbb{E}_{x}\bigl[D_{\mathrm{KL}}\bigl(p(y|x) \,\|\, p(y)\bigr)\bigr] \Bigr),
\end{equation*}
where \(p(y|x)\) is the conditional probability distribution of labels given an image \(x\) as predicted by a pretrained Inception network, and \(p(y)\) is the marginal distribution of the overall generated images. A higher IS indicates that the generated images are both diverse and of high quality.

Collectively, these metrics offer a robust evaluation framework: lower FID and KID scores signal better performance, while a higher IS indicates better generative quality.

\section{Binary Classification Results}\label{binary_results}
In this section, we evaluate the performance of the four experimental binary-class configurations, where the model is trained to generate images from a single class (birds). We report both quantitative findings (loss curves, FID, KID, and IS) and qualitative insights based on the generated images.
\subsection{Quantitative Evaluation}\label{sec:quantitative_ev}
Our Quantitative evaluation consists of loss and metrics analysis. In the loss analysis, we examine the generator and discriminator loss curves over 100 epochs. In the metric analysis, we compare the four experiments using FID, KID, and IS to quantify the fidelity and diversity of the images.

\subsubsection{Loss Results}
\begin{figure*}[!h]
	\centering
	\subfloat[Losses Plot of Experiment 1 (Mean $\pm$ Std)] {\includegraphics[width=0.45\textwidth]{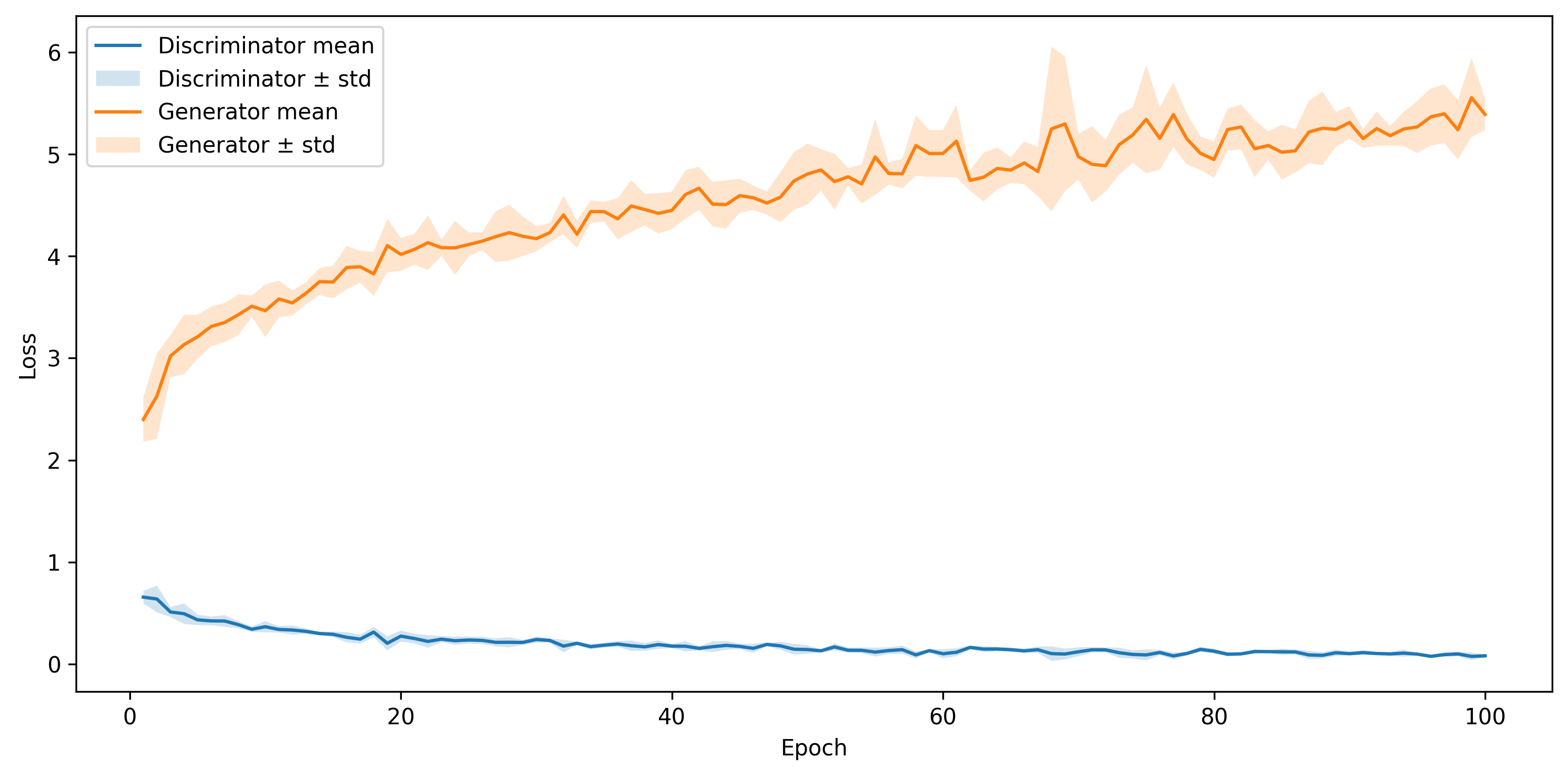}
		\label{fig:exp1Loss}
	}
	\hfill
	\subfloat[Losses Plot of Experiment 2 (Mean $\pm$ Std)] {\includegraphics[width=0.45\textwidth]{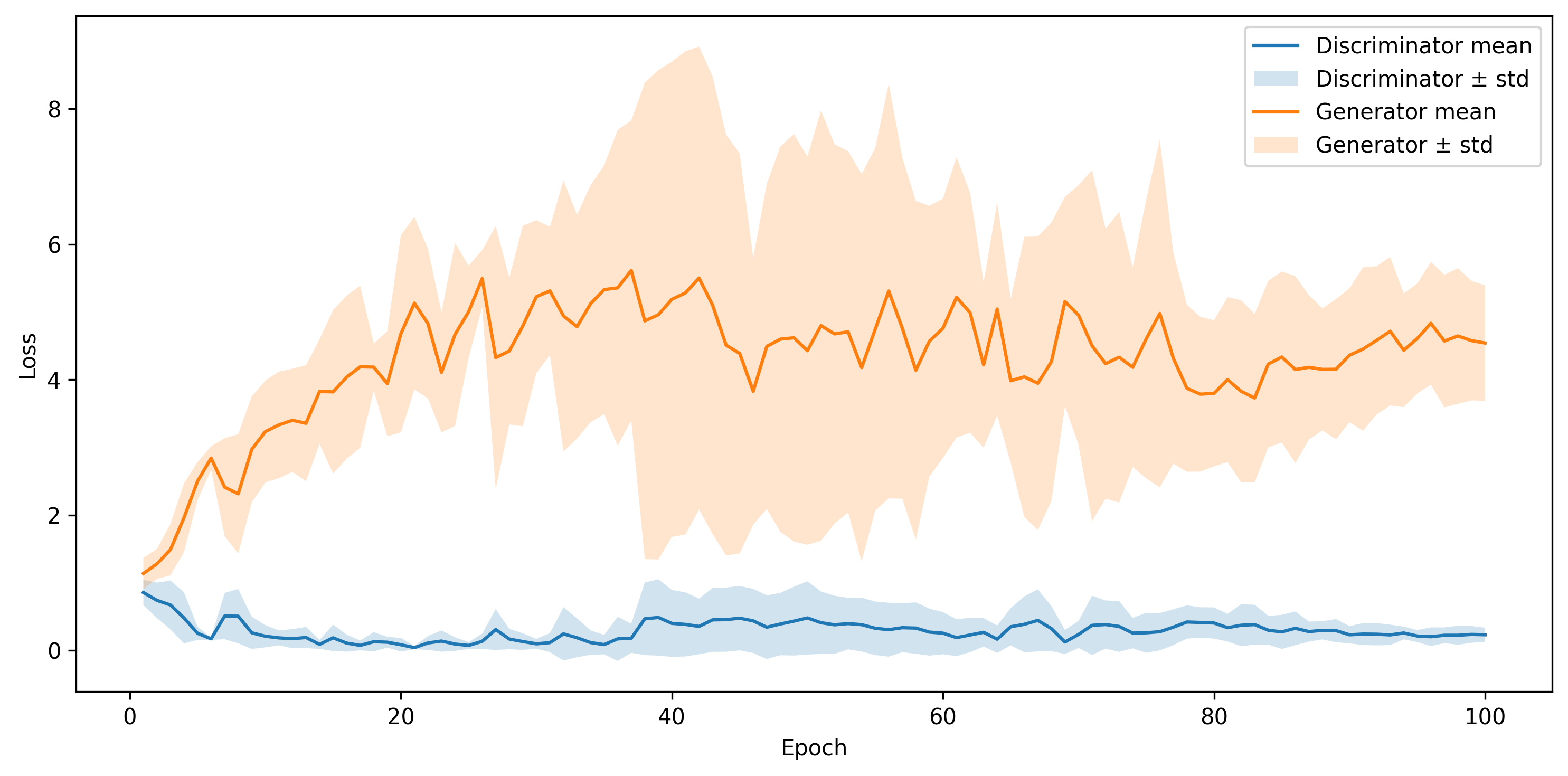}
		\label{fig:exp2Loss}
	}
        \hfill
        \subfloat[Losses Plot of Experiment 3 (Mean $\pm$ Std)] {\includegraphics[width=0.45\textwidth]{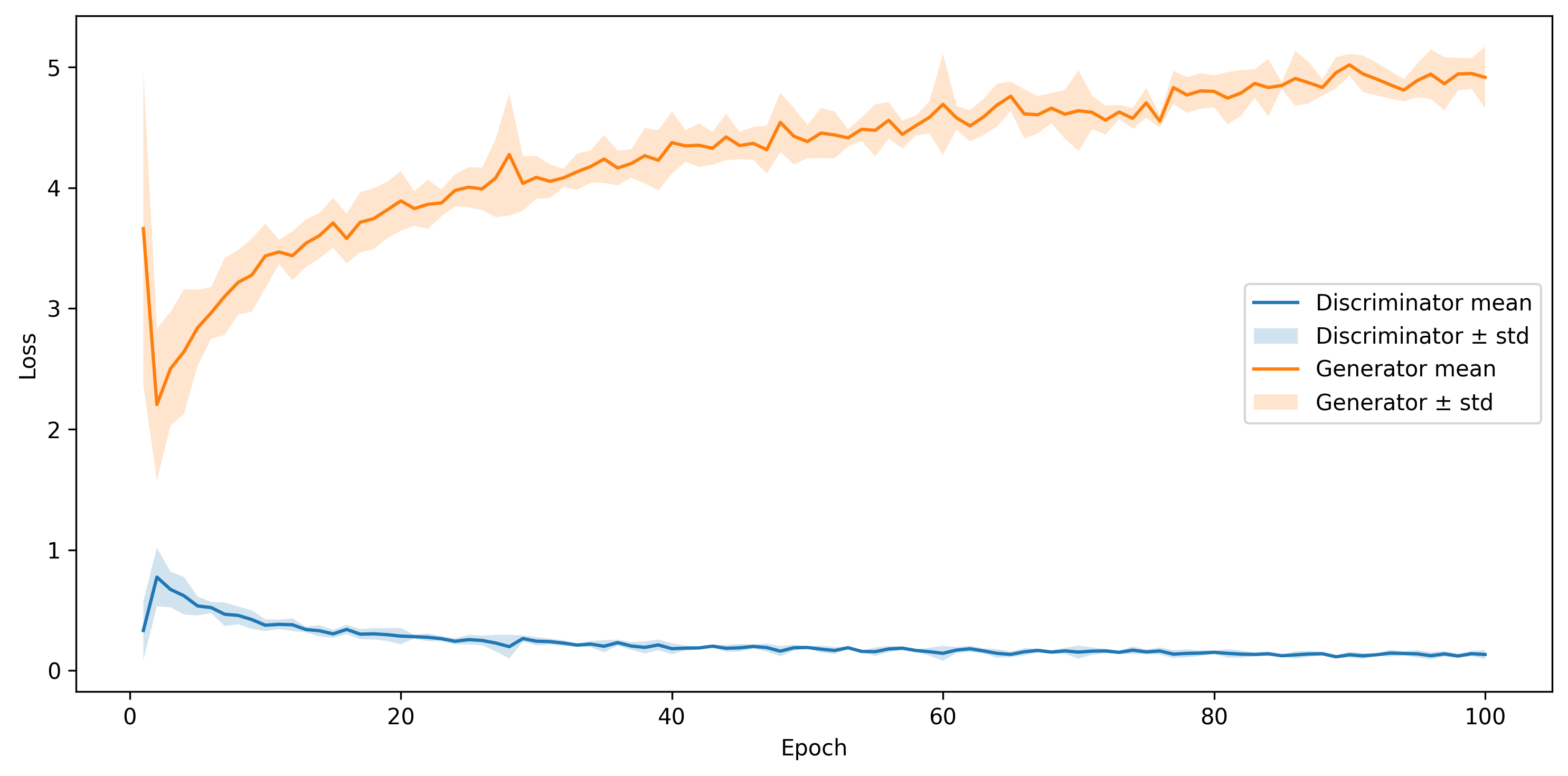}
		\label{fig:exp3Loss}
	}
	\hfill
	\subfloat[Losses Plot of Experiment 4 (Mean $\pm$ Std)] {\includegraphics[width=0.45\textwidth]{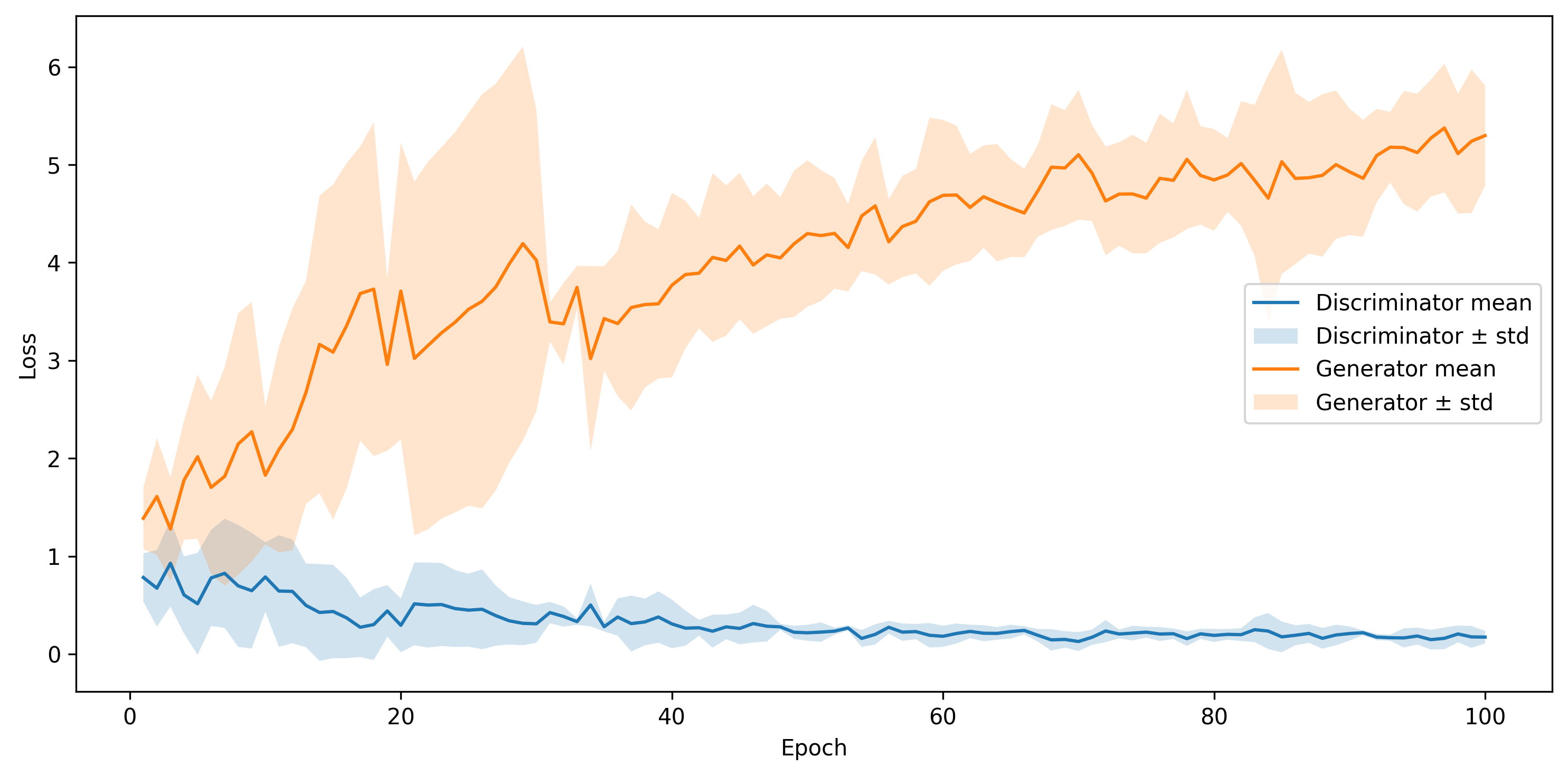}
		\label{fig:exp4Loss}
	}
	\caption{Mean Training Loss Curves with Standard Deviation Shaded Regions of Experiments 1–4}
	\label{Loss-plots}
\end{figure*}

Figures \ref{fig:exp1Loss} through \ref{fig:exp4Loss} present the generator (orange) and discriminator (blue) mean loss curves, aggregated over four independent runs for each of the four experimental configurations. The shaded regions represent the standard deviation, providing insight into the
consistency and stability of the training dynamics across different random initializations. By comparing these curves, we can observe distinct patterns of convergence, training stability, and adversarial interaction in each model.

\paragraph{Experiment 1: Classical Generator with Classical Discriminator} In this baseline model, the discriminator's loss starts at approximately 0.7 and gradually declines before stabilizing in the lower loss range. Concurrently, the generator's loss rises steadily from approximately 2.4 to the 5.0--5.5 range. This interplay suggests that the classical discriminator becomes increasingly effective at separating real and fake samples during training. The relatively narrow standard deviation bands across epochs indicate that this behavior is consistent across runs. While the classical generator avoids immediate collapse, the high terminal loss suggests that it struggles to produce samples that can effectively challenge the discriminator as training progresses.

\paragraph{Experiment 2: Classical Generator with Hybrid Discriminator.} In this experiment, only the discriminator incorporates a quantum block. The discriminator's loss drops more sharply in the early epochs (from roughly 0.8 to very low levels within the first 15--20 epochs) before stabilizing at very low levels, reflecting a heightened capacity to distinguish real from fake images. The generator's loss, meanwhile, rises rapidly and fluctuates throughout
training. The comparatively wide standard deviation region in this experiment indicates greater variability across runs, suggesting that the quantum-enhanced discriminator provides a stronger adversary, pushing the classical generator to more volatile loss values.

\paragraph{Experiment 3: Hybrid Generator with Classical Discriminator} When the generator is augmented with a quantum block, but the discriminator remains classical, the discriminator’s loss shows a brief early rise before gradually settling to a low but nonzero range. Compared with Experiment 1, this indicates a short initial re-balancing phase in the adversarial interaction, after which the training dynamics become comparatively stable. The relatively narrow standard deviation bands indicate stable training behavior across runs, while the absence of the stronger fluctuations seen in Experiment 2 indicates that placing the quantum block in the generator yields smoother training than placing it only in the discriminator.

\paragraph{Experiment 4: Hybrid Generator with Hybrid Discriminator} In the fully hybrid model, both networks incorporate quantum blocks. The discriminator's loss drops rapidly in the early epochs, mirroring some patterns seen in Experiment 2, but stabilizes with moderate fluctuations across training. The generator's loss rises rapidly in the early stage, with noticeable volatility, before settling into a moderately high range, indicating sustained adversarial tension. Compared to Experiment 2, the standard deviation region is somewhat narrower during later epochs, indicating more stable behavior across runs. The overall pattern suggests that both networks are leveraging quantum features, leading to a more sustained yet intense adversarial interaction, although the discriminator remains the stronger side for much of training.

Taken as a whole, these loss plots suggest that VQC placement meaningfully affects training behavior. Experiment 2, where the quantum block is introduced only in the discriminator, shows the greatest variability and the widest standard deviation bands, indicating less consistent dynamics across runs. By contrast, Experiments 1 and 3 remain comparatively tighter, while Experiment 4 appears more controlled than Experiment 2 in later epochs, despite some early volatility. These patterns suggest that placing quantum components in both networks may support a more sustained adversarial interaction than using a quantum discriminator alone. While these loss trends are informative about training stability, their implications for generative quality are interpreted together with the quantitative metrics and qualitative results presented in the following subsections.

\subsubsection{Metrics Results}%\vspace{-1.0em}
Table~\ref{tab:extended_metrics} summarizes the FID, KID, and IS across four independent runs for each experiment. Lower FID and KID values indicate that the generated images are closer to the real-image distribution, whereas a higher IS implies better quality and diversity in the generated samples.

\begin{table*}[!h]
\centering
\begin{tabular}{|l|l|l|l|l|l|}
\hline
\textbf{Ex} & \textbf{Generator} & \textbf{Discriminator} & \textbf{FID ($\downarrow$)} & \textbf{KID ($\downarrow$)} & \textbf{IS ($\uparrow$)} \\ \hline
1 & Classical & Classical & 321.9 $\pm$34.2 & 0.27 $\pm$0.09 & 1.85 $\pm$0.06 \\ \hline
2 & Classical & Hybrid & 319.4 $\pm$17.9 & 0.23 $\pm$0.02 & 2.26 $\pm$0.12 \\ \hline
3 & Hybrid & Classical & 250.0 $\pm$15.5 & 0.14 $\pm$0.02 & 2.54 $\pm$0.34 \\ \hline
4 & Hybrid & Hybrid & \textbf{218.5 $\pm$16.7} & \textbf{0.11 $\pm$0.02} & \textbf{2.55 $\pm$0.44} \\ \hline
\end{tabular}
\caption{Quantitative Performance Metrics (FID, KID, and IS) Across Four Independent Runs (Mean $\pm$ Std)}
\label{tab:extended_metrics}
\end{table*}

The baseline model (Experiment 1) exhibits the highest FID (321.9 $\pm$ 34.2) and KID (0.27 $\pm$ 0.09), along with the lowest IS (1.85 $\pm$ 0.06). This fully classical setup struggles to match the performance of the hybrid models. By introducing a quantum block into the discriminator (Experiment 2), FID decreases to 319.4 $\pm$ 17.9 and KID to 0.23 $\pm$ 0.02, while IS increases to 2.26 $\pm$ 0.12. Although this outperforms Experiment 1, it does not represent the best overall outcome. When only the generator is quantum-enhanced (Experiment 3), FID reduces to 250.0 $\pm$ 15.5, KID to 0.14 $\pm$ 0.02, and IS increasing to 2.54 $\pm$ 0.34. This indicates a substantial improvement in quantitative performance relative to Experiments 1 and 2. In the fully hybrid configuration (Experiment 4), FID and KID reach their lowest mean values (218.5 $\pm$ 16.7 and 0.11 $\pm$ 0.02, respectively), indicating the strongest distributional similarity to the real data, while IS remains the highest at 2.55 $\pm$ 0.44.

Compared with the classical baseline, Experiment 2 achieves a 0.78\% reduction in FID, a 14.81\% reduction in KID, and a 22.16\% increase in IS. Experiment 3 shows substantially stronger gains, with a 22.34\% reduction in FID, a 48.15\% reduction in KID, and a 37.30\% increase in IS. The fully hybrid model in Experiment 4 delivers the strongest overall improvement, achieving a 32.12\% reduction in FID, a 59.26\% reduction in KID, and a 37.84\% increase in IS, with the reported standard deviations indicating that these improvements are consistent across repeated runs.

    \begin{figure*}[!h]
      \centering
      \includegraphics[width=1\textwidth]{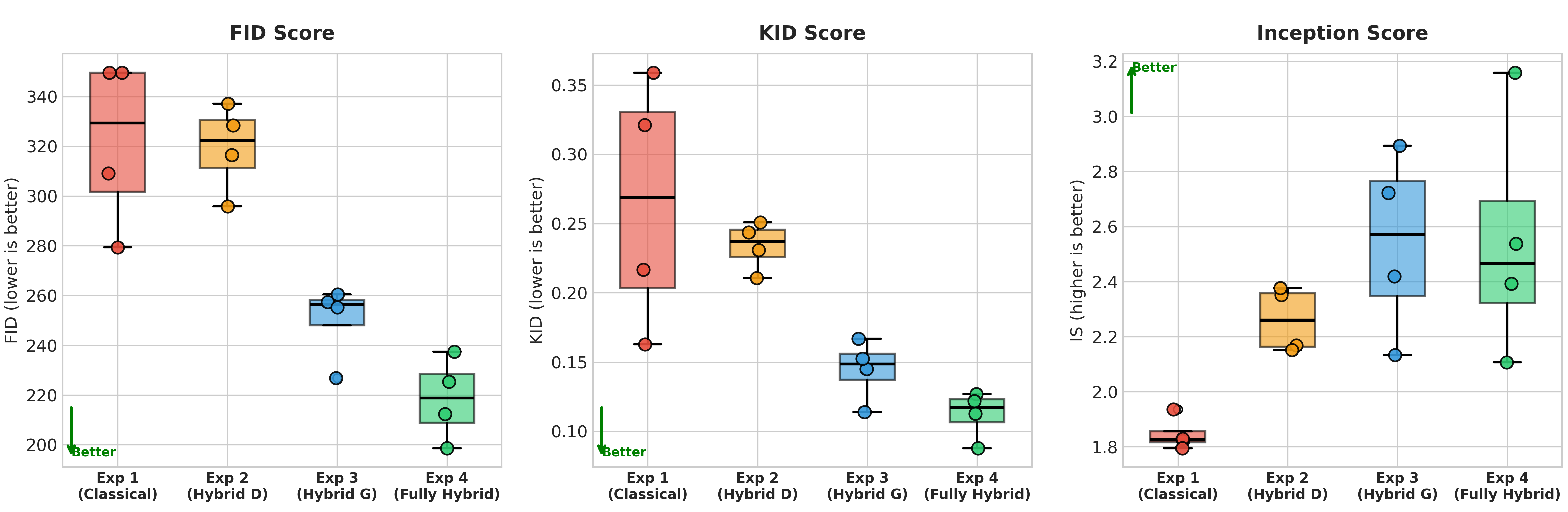}
      	\hspace{0.5em}
        \caption{Distribution of Quantitative Performance Metrics (FID, KID, and IS) Across Four Independent Runs for Each Experimental Configuration}
      \label{fig:boxplot} 
    \end{figure*}
    
Figure~\ref{fig:boxplot} presents the distribution of FID, KID, and IS scores across four independent runs for each experimental configuration. Each data point corresponds to an individual run, and the box plots summarize the median, interquartile range, and full extent of the results. For FID, the distributions shift progressively lower from Experiment 1 through Experiment 4, with the fully hybrid model achieving the lowest scores across all runs. A similar trend is observed for KID, where Experiment 4 exhibits both the lowest median and the smallest spread, indicating consistent performance across different initializations. For IS, all hybrid configurations outperform the classical baseline, though the variance increases in Experiments 3 and 4. Across all three metrics, the distribution ranges of Experiment 4 remain separated from those of Experiment 1, confirming that the observed differences are consistent across runs.

%\paragraph{Metric Biases and Limitations}\leavevmode\par
However, it is important to acknowledge the biases and limitations of certain evaluation metrics. As noted by \cite{Alhamdi2024}, FID can be biased toward specific datasets, given its reliance on feature embeddings from a pre-trained inception network. Similarly, IS has been criticized for its unreliability in capturing the actual realism of the generated samples. Therefore, relying on additional metrics such as KID, which is less prone to domain-specific biases, is crucial for a more comprehensive evaluation of generative performance.

\subsection{Qualitative Evaluation}\label{sec:qualitative_ev}
\begin{figure*}[!h]
	\centering
	\subfloat[Generated Samples from Experiment 1] {\includegraphics[width=0.45\textwidth]{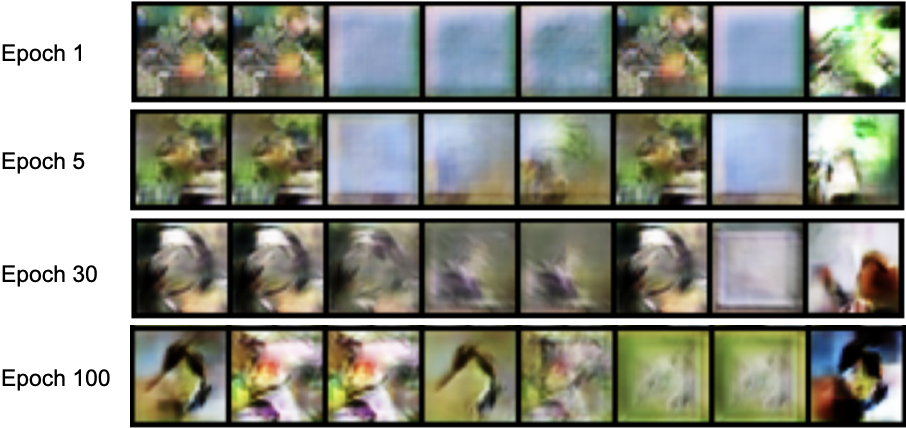}
		\label{fig:exp1Gen}
	}
	\hfill
	\subfloat[Generated Samples from Experiment 2] {\includegraphics[width=0.45\textwidth]{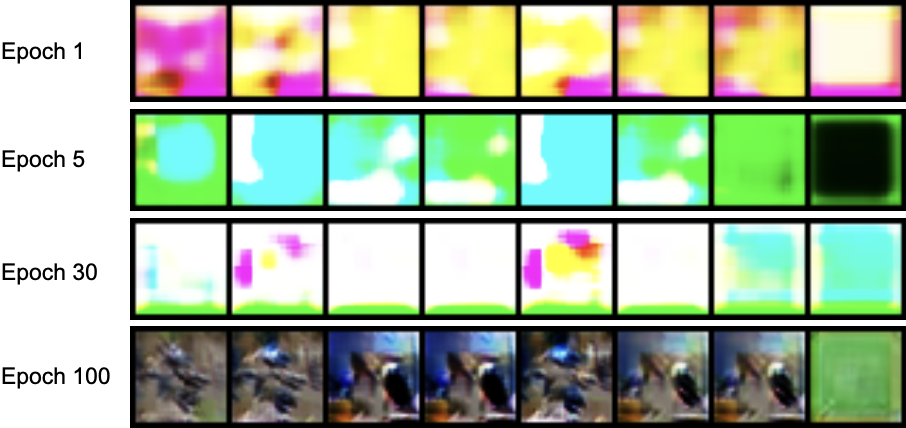}
		\label{fig:exp2Gen}
	}
        \hfill
        \subfloat[Generated Samples from Experiment 3] {\includegraphics[width=0.45\textwidth]{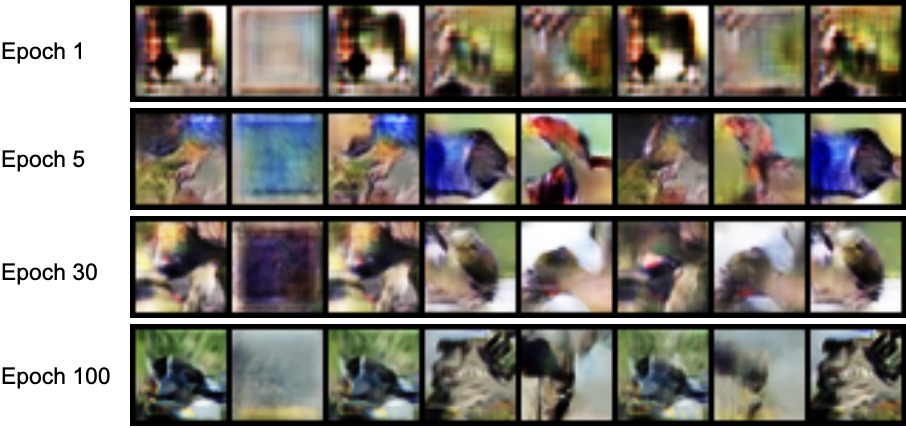}
		\label{fig:exp3Gen}
	}
	\hfill
	\subfloat[Generated Samples from Experiment 4] {\includegraphics[width=0.45\textwidth]{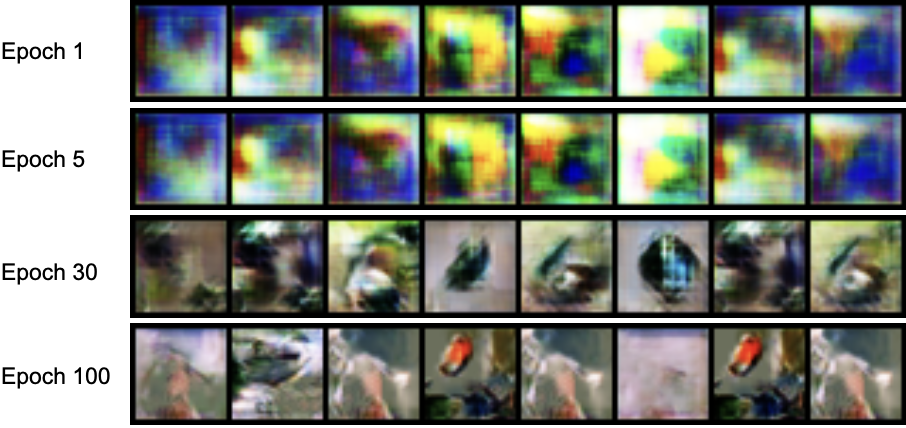}
		\label{fig:exp4Gen}
	}
	\caption{Generated Samples from Experiments 1 - 4}
	\label{Generated}
\end{figure*}

One key observation that we can conclude from visually analyzing the generated images is that when the discriminator includes a quantum block but the generator remains classical (as in Experiment~2, Figure~\ref{fig:exp2Gen}), the model appears to converge slower in the early epochs -- reflected by the highly abstract and less-defined outputs at epochs~1,~5, and~30. In contrast, placing the quantum block in the generator (Experiment~3, Figure~\ref{fig:exp3Gen}) seems to accelerate the initial learning process; even within the first few epochs, the model begins to form rudimentary bird outlines, suggesting that the quantum-enabled generator rapidly captures essential structural information compared to the other models. 
In the fully hybrid setup (Experiment~4, Figure~\ref{fig:exp4Gen}), the generator's quantum capacity competes with the quantum-enhanced discriminator, which again introduces some early-stage volatility but ultimately leads to more refined images.  This qualitative observation is grounded by the quantitative results in Table~\ref{tab:extended_metrics}, where the fully hybrid model achieves the strongest overall performance, with a 32.12\% reduction in FID, a 59.26\% reduction in KID, and a 37.84\% increase in IS relative to the classical baseline. Overall, these observations reinforce that the location of the quantum block (generator vs.\ discriminator) may influence the speed and character of convergence during the initial phases of training.

%%%%%%%%%%%%%%%%%%%%%%%Generated Images

%%%%%%%%%%%%%%%%%%%%%%%%%%%%%%%%%%%%%%%%%
\section{Multi-Class Classification Results}\label{multiclass_results}
In this section, we present the results of Experiment 5, where we evaluate our model on a multi-class classification problem. Unlike the binary classification experiments we conducted in Section \ref{binary_results}, in this section, we focus on the fully hybrid variant of our model as it is the one that outperformed all other variants in the binary classification problem. 
\subsection{Quantitative Evaluation}%\vspace{-1.0em}
\begin{figure*}[!h]
	\centering
	\subfloat[FID Score (Lower values are better)] {\includegraphics[width=0.3\textwidth]{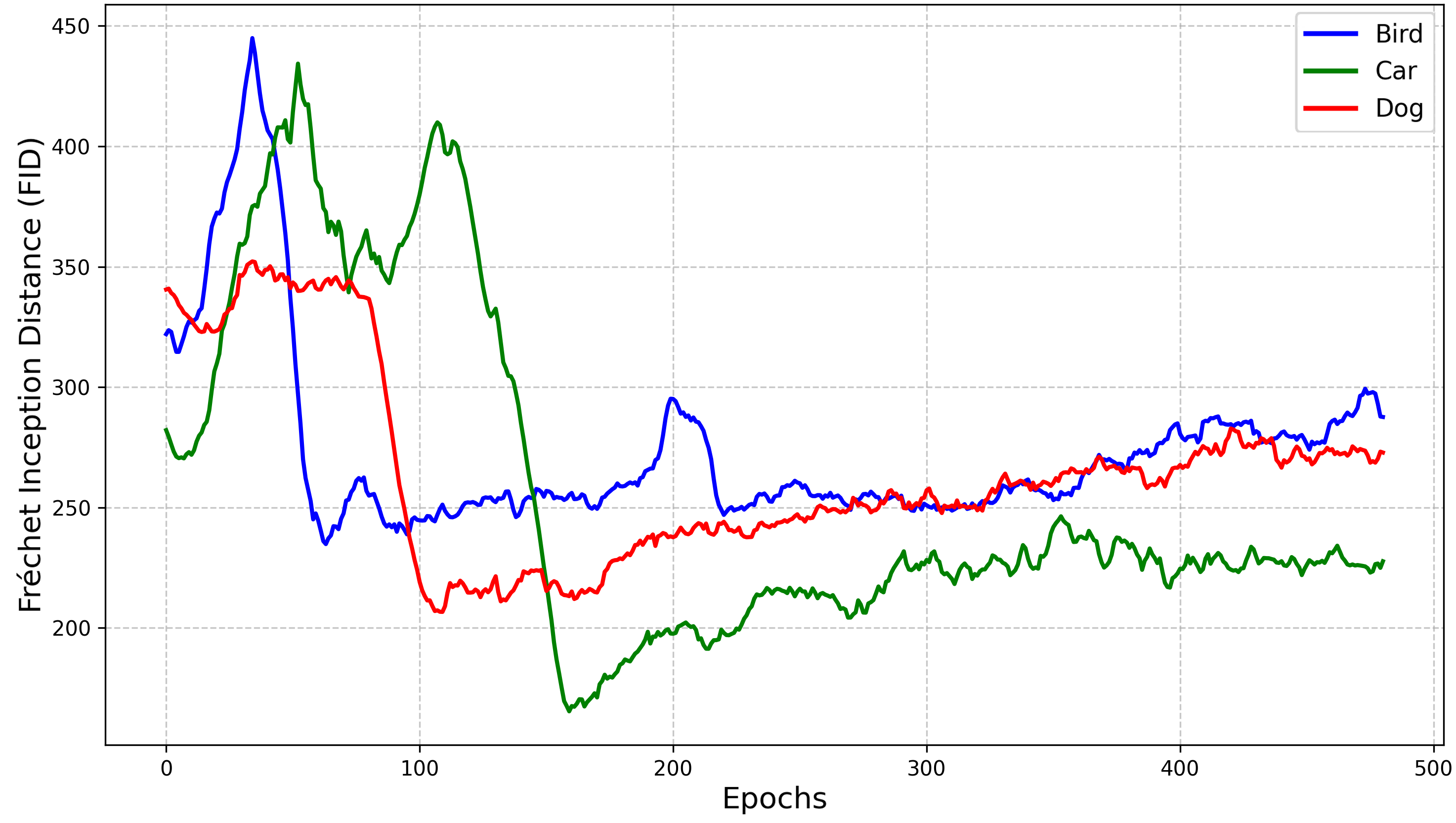}
		\label{fig:FID_5000}
	}
	\hfill
	\subfloat[KID Score (Lower values are better)] {\includegraphics[width=0.3\textwidth]{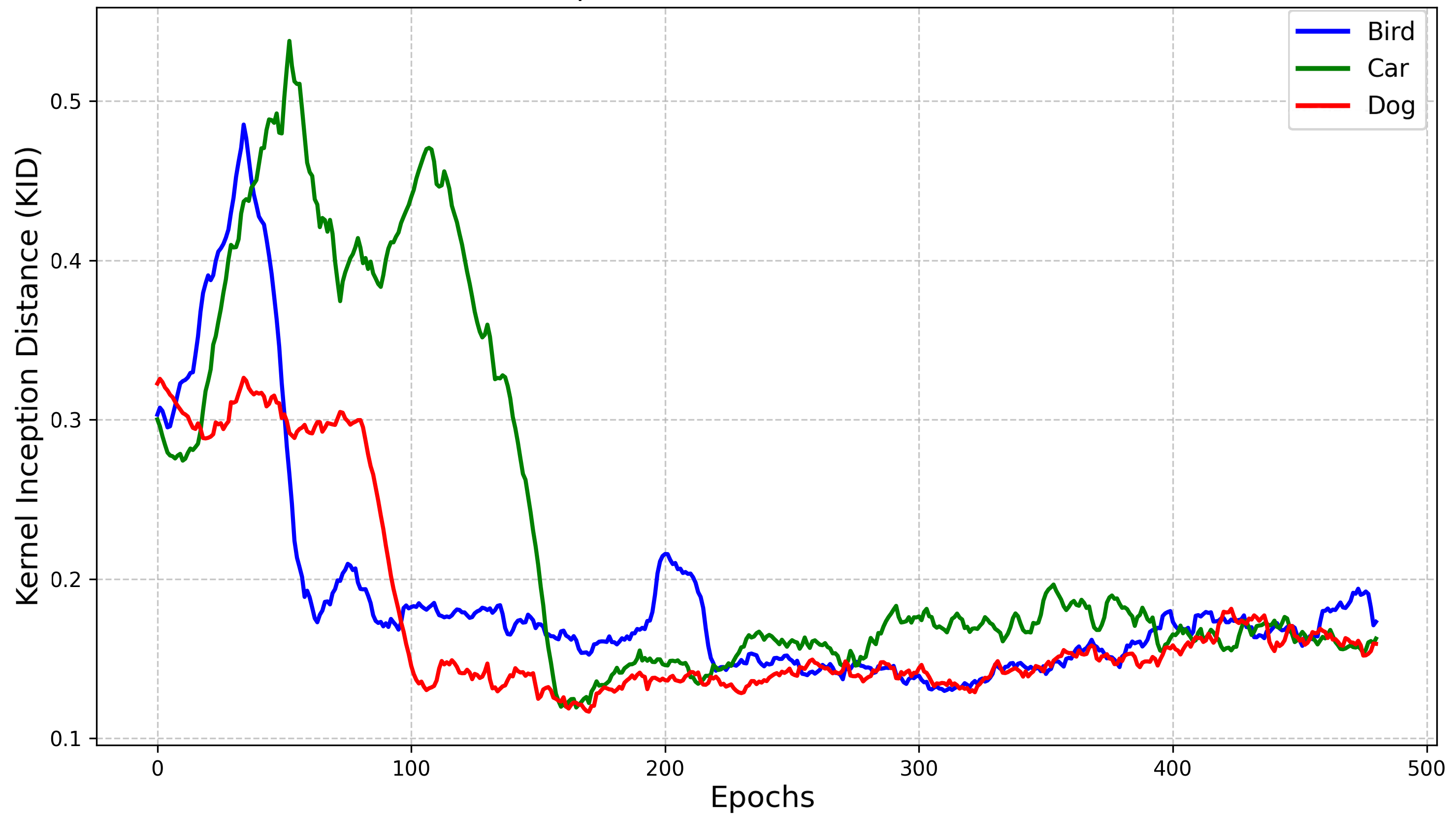}
		\label{fig:KID_5000}
	}
        \hfill
        \subfloat[IS Score (Higher values are better)] {\includegraphics[width=0.3\textwidth]{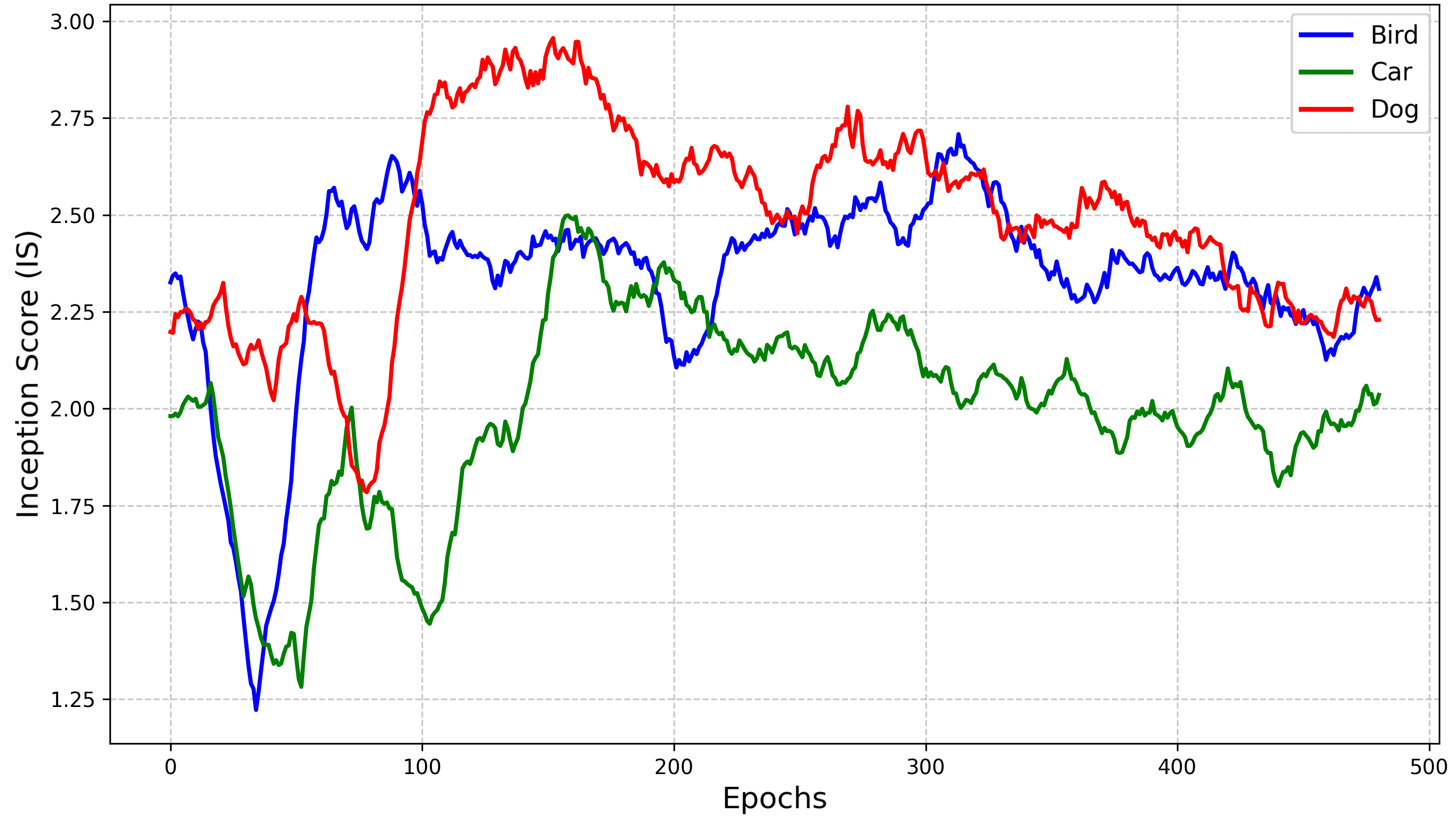}
		\label{fig:IS_5000}
	}
	\caption{FID, KID and IS Scores for Multiclass Classification with 5000 Samples}
	\label{Metrics_experiment5_5000}
\end{figure*}

    When trained on multiple classes (Birds, Cars, Dogs) in a fully hybrid setting (Experiment 5), the model typically enters a phase of initial instability followed by strong improvement during the first 100–200 epochs, showing noticeable gains in quality and distributional fidelity, as evident in Figures~\ref{fig:FID_5000}, \ref{fig:KID_5000}, and \ref{fig:IS_5000}. Beyond 200 epochs, the rate of improvement gradually slows down, most curves begin to flatten, and training enters a phase of incremental changes rather than dramatic gains. However, the later stage is not fully stable, as the metrics continue to display oscillations and, for some classes, mild regressions rather than monotonic improvement. For example, Car achieves its strongest FID/KID improvements around the mid-training stage, Dog reaches its highest IS in the mid-epoch range, and Bird shows comparatively less consistent late-stage behavior.
        %%%%%%%%%%%%%%%%%%%%%%%FID/KID/IS Plots

     %%%%%%%%%%%%%%%%%%%%%%%FID/KID/IS Plots
     
While the hybrid architecture can handle complex, multi-class data effectively, the key takeaway is that many of the most substantial improvements emerge within the first few hundred epochs and a plateau in progress (supported by the relative stabilization of the scores and qualitative inspection of the generated samples shown in Figure \ref{fig:Generated_experiment5_5000}), generally suggest the model approaches a practically stable regime and is near convergence.

\subsection{Qualitative Evaluation}%\vspace{-1.0em}
\begin{figure*}[!h]
	\centering
	\subfloat[Birds class] {\includegraphics[width=0.3\textwidth]{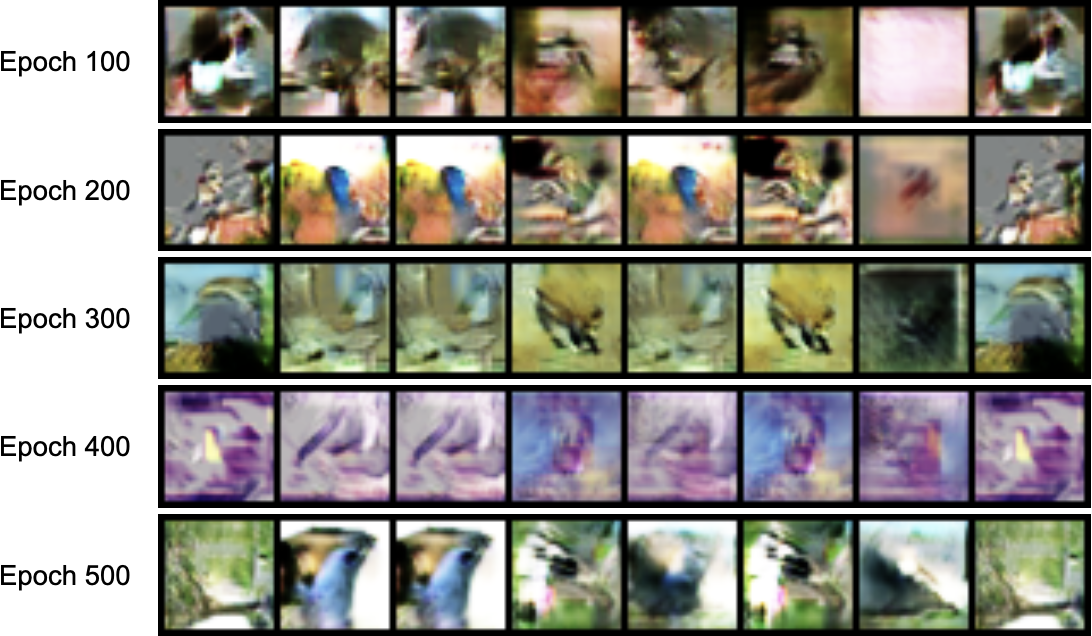}
		\label{fig:exp5_Birds}
	}
	\hfill
	\subfloat[Dogs class] {\includegraphics[width=0.3\textwidth]{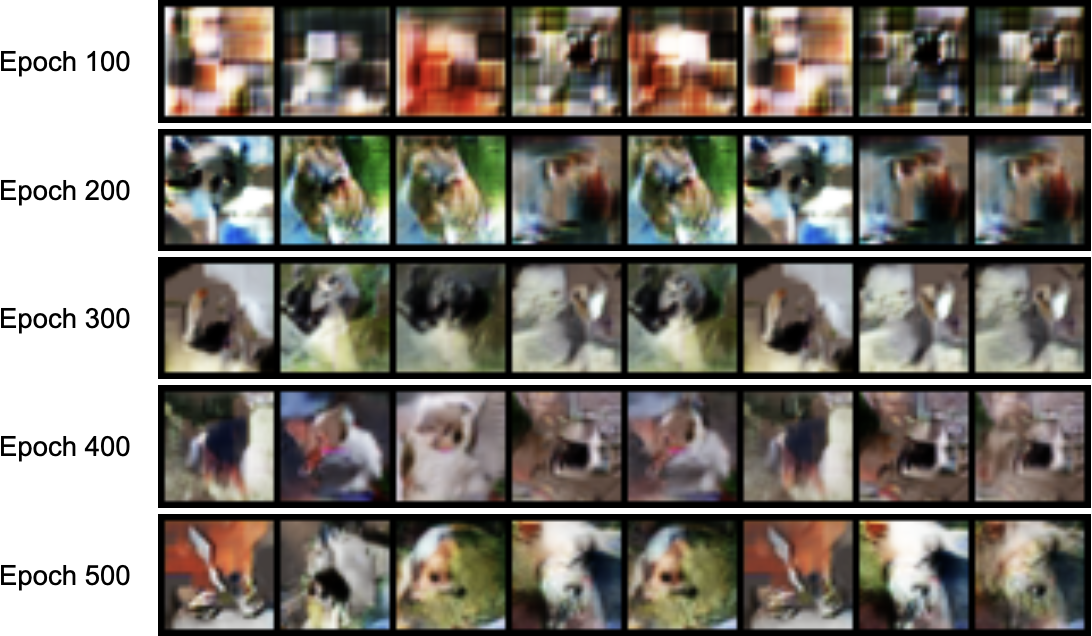}
		\label{fig:exp5_Dogs}
	}
        \hfill
        \subfloat[Cars class] {\includegraphics[width=0.3\textwidth]{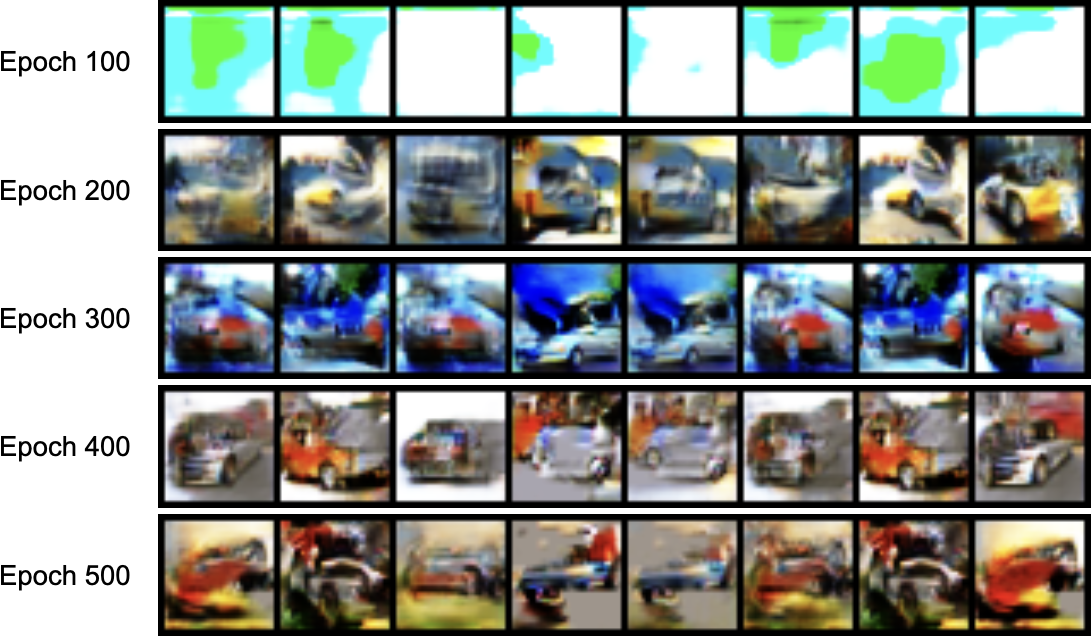}
		\label{fig:exp5_Cars}
	}
	\caption{Generated Samples from Experiment 5 with 5000 samples}
	\label{fig:Generated_experiment5_5000}
\end{figure*}

Figures \ref{fig:exp5_Birds}, \ref{fig:exp5_Cars}, and \ref{fig:exp5_Dogs} illustrate the progression of generated samples at different training stages (epochs 100, 200, 300, 400, and 500). These visualizations confirm that most substantial improvements occur within the first 200 epochs, after which refinements are more gradual, supporting the earlier observation that the model reaches near-convergence around epoch 200.

\section{Learning Curve Results}\label{learning_curves}
\begin{figure*}[!h]
	\centering
	\subfloat[FID Score (Lower values are better)] {\includegraphics[width=0.3\textwidth]{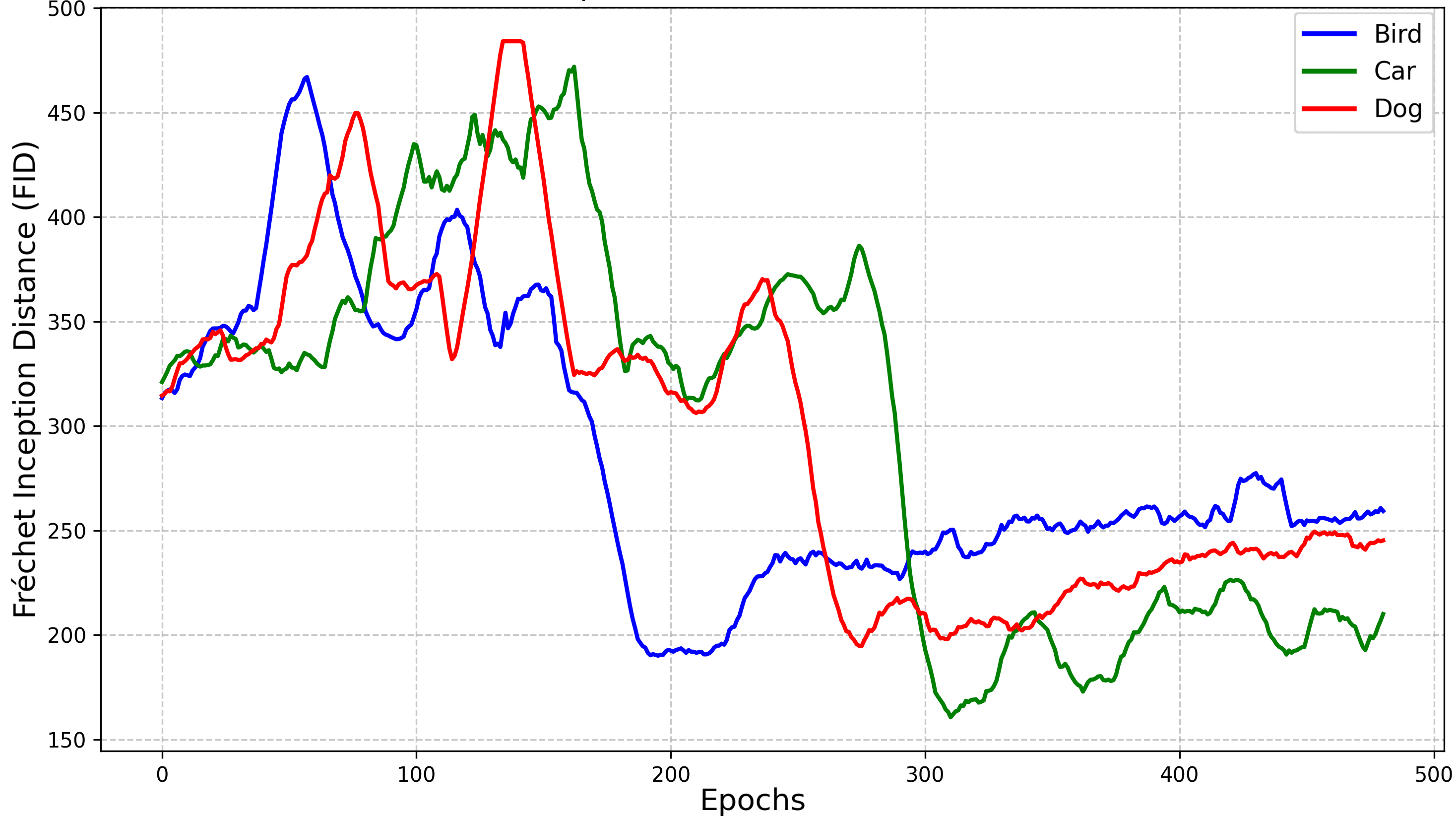}
		\label{fig:FID_2500}
	}
	\hfill
	\subfloat[KID Score (Lower values are better)] {\includegraphics[width=0.3\textwidth]{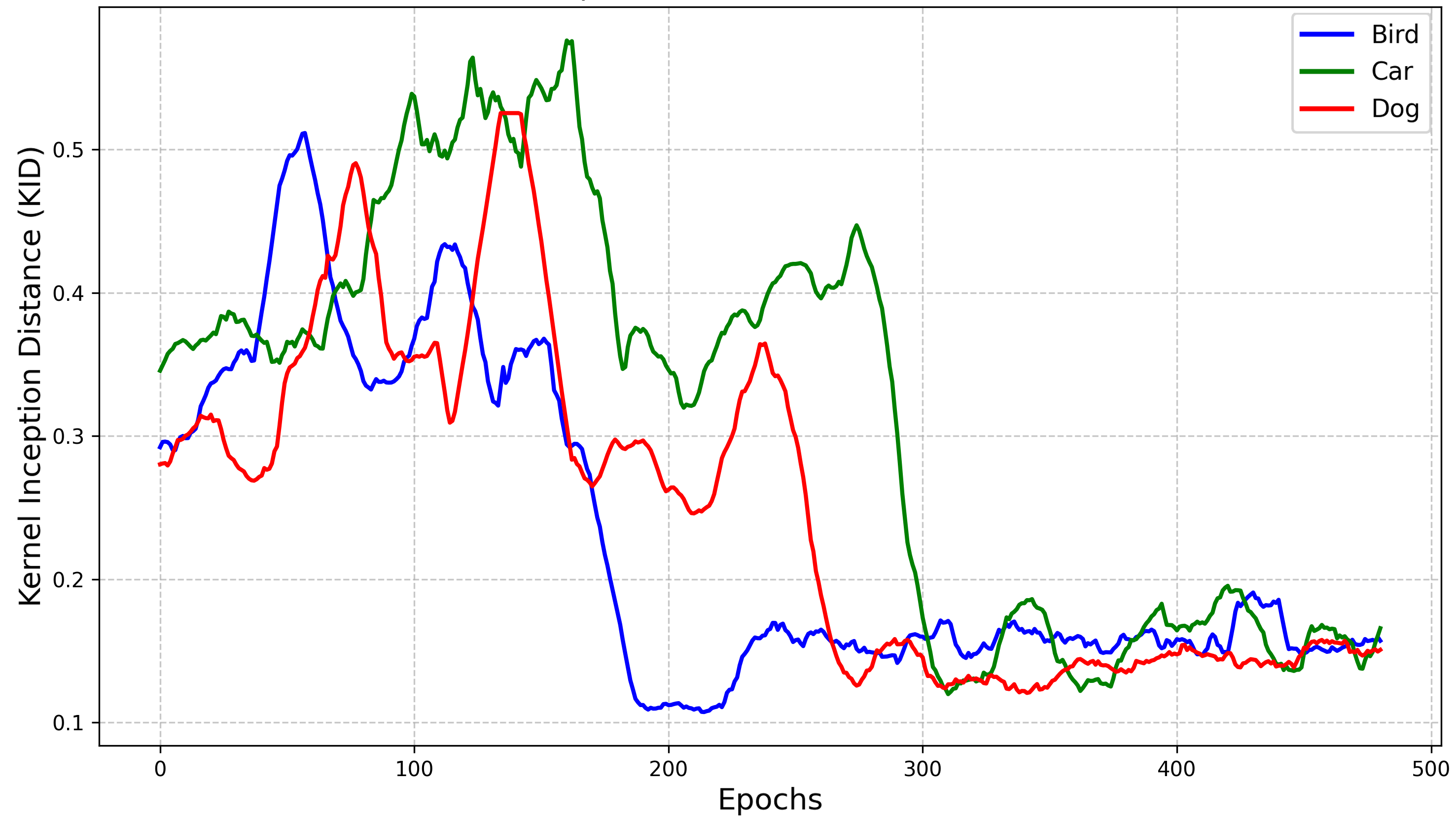}
		\label{fig:KID_2500}
	}
        \hfill
        \subfloat[IS Score (Higher values are better)] {\includegraphics[width=0.3\textwidth]{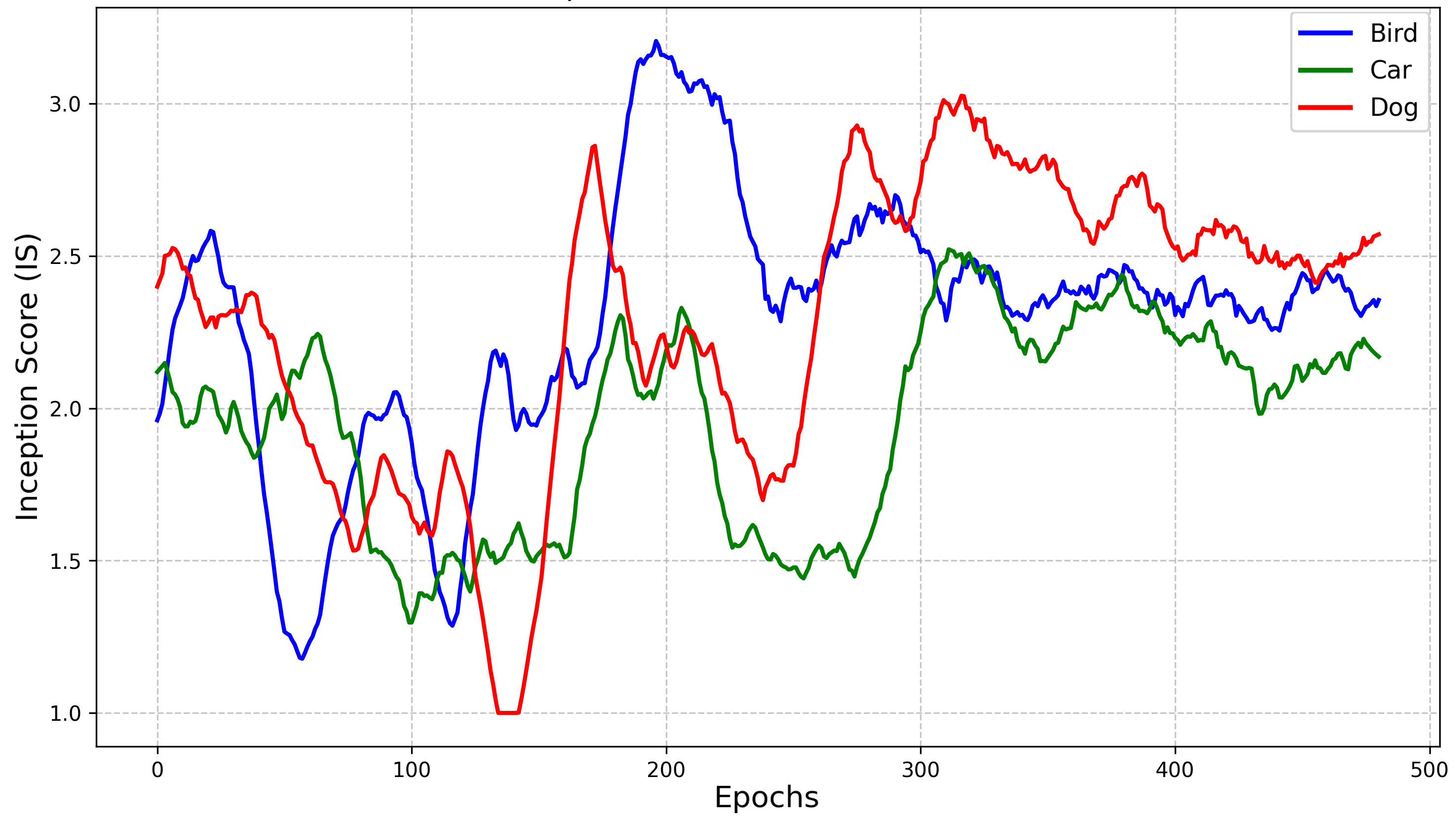}
		\label{fig:IS_2500}
	}
	\caption{FID, KID and IS Scores for Multiclass Classification with 2500 Samples}
	\label{Metrics_experiment5_2500}
\end{figure*}

When the sample size is reduced to 2,500 per class, the early training stages (epoch 0-200) show increased volatility and slower convergence compared to the 5,000-sample runs. However, as training progresses, the key metrics, such as FID (Figure~\ref{fig:FID_2500}), KID (Figure~\ref{fig:KID_2500}), and IS (Figure~\ref{fig:IS_2500}), eventually converge to levels that are comparable to those achieved with a larger sample size in Figures \ref{fig:FID_5000}, \ref{fig:KID_5000}, and \ref{fig:IS_5000}.
        %%%%%%%%%%%%%%%%%%%%%%%FID/KID/IS Plots

\begin{figure*}[!h]
	\centering
	\subfloat[Birds class] {\includegraphics[width=0.3\textwidth]{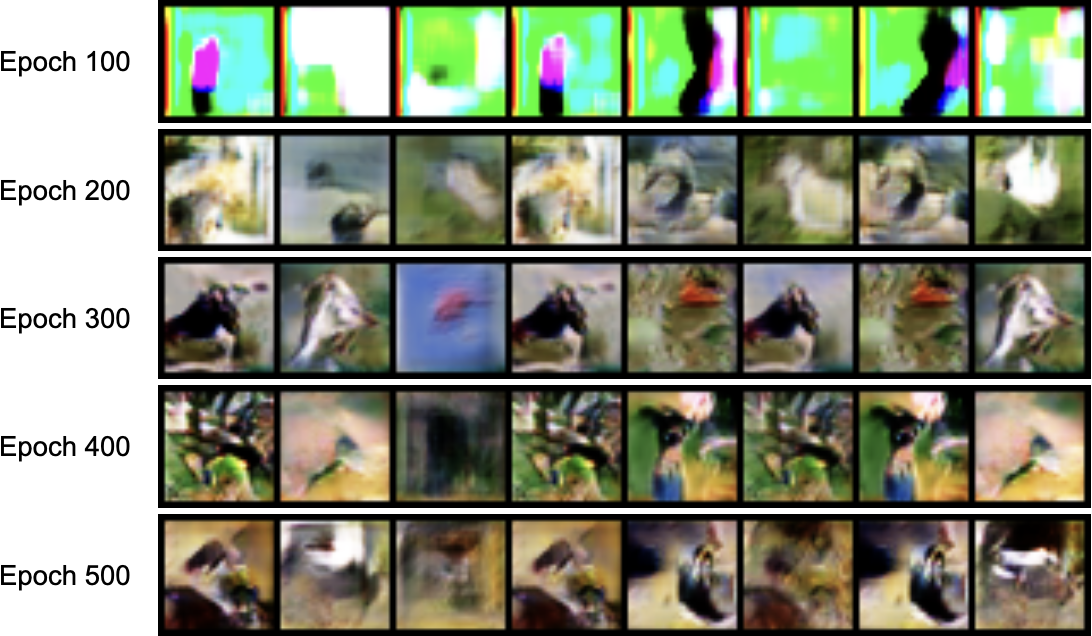}
		\label{fig:exp5_Birds_2500}
	}
	\hfill
	\subfloat[Dogs class] {\includegraphics[width=0.3\textwidth]{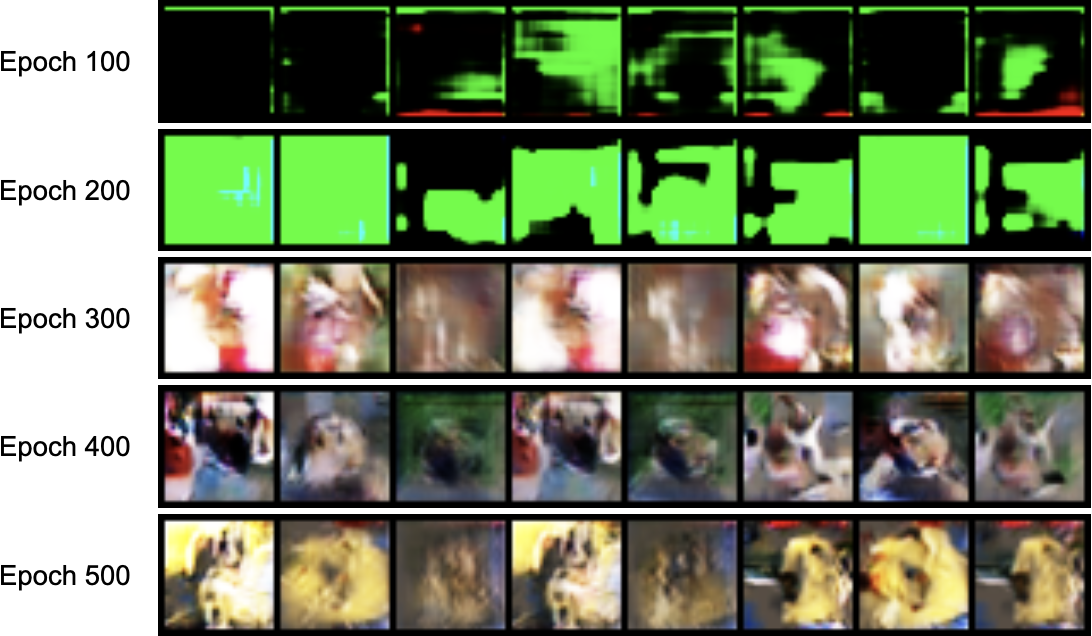}
		\label{fig:exp5_Dogs_2500}
	}
        \hfill
        \subfloat[Cars class] {\includegraphics[width=0.3\textwidth]{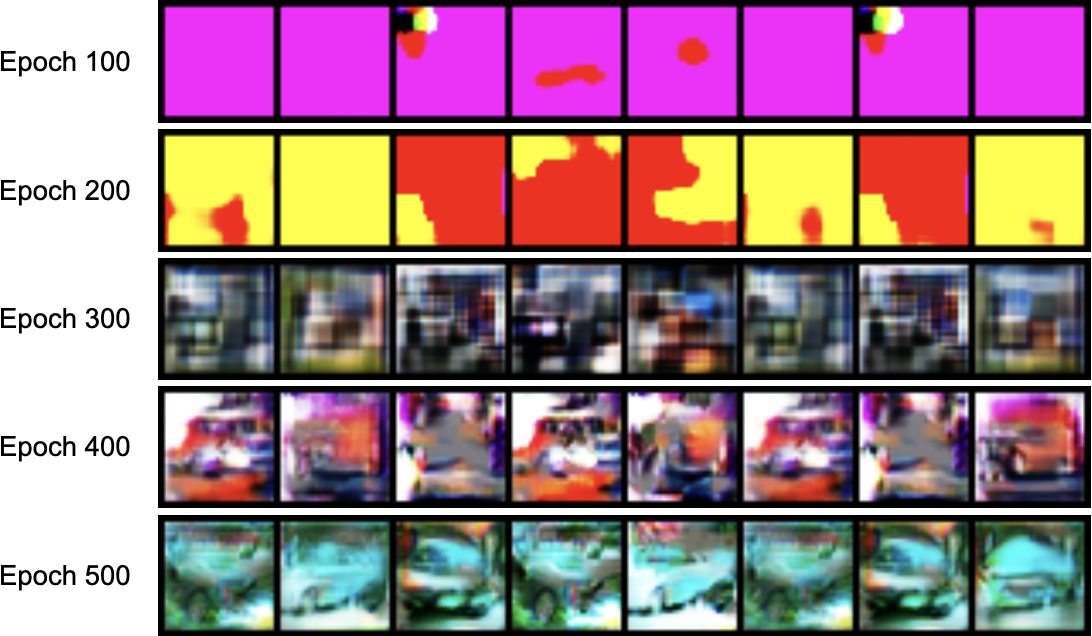}
		\label{fig:exp5_Cars_2500}
	}
	\caption{Generated Samples from Experiment 5 with 2500 samples}
	\label{Generated_experiment5_2500}
\end{figure*}

To further validate the observed stability in the 2500-sample setting, Figures \ref{fig:exp5_Birds_2500}, \ref{fig:exp5_Dogs_2500}, and \ref{fig:exp5_Cars_2500} present the qualitative evolution of generated images across training epochs. Despite an initially noisier training process, the generator progressively refines object structures and textures, producing outputs comparable to those from the larger dataset.

This suggests that despite the initial instability, the model is able to capture the essential features required for image synthesis even with fewer samples. These findings suggest that the fully hybrid design remains effective even under data-limited conditions, and indicate that the transfer learning backbone remains practically beneficial within the more demanding settings.

\section{Conclusion}\label{conclusion}
In this paper, we explored the integration of VQCs and transfer learning techniques into GANs. We adopted an architectural perspective and asked where a quantum layer should be placed within the GAN in order to be most effective. To this end, we fixed the VQC design and systematically compared fully classical and hybrid configurations in which the same VQC is inserted into the generator, the discriminator, or both. Through our experimental evaluation, we assessed the performance of various configurations across multiple classes of the CIFAR-10 dataset and training conditions, showing that, in our setting, hybrid quantum-classical GANs can achieve improved quantitative metrics compared to a purely classical baseline. The significance of this work lies in demonstrating how the placement of quantum layers within a GAN influences convergence behavior, representation capacity, and sample fidelity. Furthermore, we leveraged transfer learning, by incorporating pre-trained feature extractors into the discriminator to support feature extraction and training stability. By bridging quantum computing and deep learning, this research contributes to the ongoing effort to leverage near-term quantum devices for practical machine learning applications. While the experiments are conducted in classical simulation, the VQC is intentionally kept small and shallow to reflect realistic NISQ constraints. Although noise on real devices would lower absolute performance, the architectural trends arising from the placement of the quantum layer may still remain valid, as they depend primarily on the functional role of the circuit rather than its depth.

Our results demonstrate that incorporating quantum blocks into the GAN architecture can improve both the qualitative and quantitative performance of the generated images in the studied setting. The fully hybrid model, which integrates quantum blocks in both the generator and the discriminator, outperforms the fully classical baseline as well as the mixed configurations. Notably, a quantum-enabled generator accelerates early visual convergence and better captures intricate structural details, while a quantum-enhanced discriminator introduces a more challenging adversarial signal, slowing early visual convergence but improving the final quantitative metrics relative to the baseline. Furthermore, even with reduced sample sizes, the fully hybrid model converges to performance levels comparable to those achieved with larger datasets, suggesting that the proposed design remains effective under reduced-data conditions. These findings suggest that quantum components can enhance the expressive power of GANs, thereby opening new avenues for future research focused on optimizing hybrid quantum-classical architectures and extending their applications to high-resolution image synthesis and other complex generative tasks. An important direction for future work is to investigate how such few-qubit, shallow VQCs could be scaled to deeper circuits and larger qubit counts, for example via patch-wise tiling strategies or more expressive hardware-efficient circuit architectures on emerging devices as well as applying these models to specialized domains, such as medical image synthesis.

\section*{Code Availability}
The source code used in this paper is available at the following GitHub repository: 
\url{ https://github.com/QC2-HBKU/Asma-HQGAN-TL}

\bibliographystyle{IEEEtran}
\bibliography{bibtex/bib/IEEEexample}

\end{document}